\documentclass[acmsmall, nonacm]{acmart}

\usepackage{booktabs} 
\usepackage{color}
\usepackage{colortbl}
\usepackage{subcaption}
\usepackage{stfloats}
\usepackage{todonotes}
\usepackage{float}
\usepackage{longfbox}
\usepackage{subcaption}
\usepackage{enumitem}

\newcommand{\eg}{\textit{e.g.}}
\newcommand{\ie}{\textit{i.e.}}
\newcommand{\cf}{\textit{c.f.}}

\definecolor{usercolor}{HTML}{eaeaea}

\newfboxstyle{patternparam}{padding=1.5pt, margin-bottom=1.5pt, margin-top=1.5pt, border-style=none, height=8pt}
\newcommand{\slotbox}[1]{\lfbox[patternparam, background-color=usercolor]{#1}}

\definecolor{jing}{RGB}{27,158,119}
\definecolor{sungdong}{RGB}{2,95,217}
\definecolor{youngho}{RGB}{102,166,30}
\definecolor{hyunhoon}{RGB}{20, 201, 192}

\definecolor{revisedcolor}{RGB}{0,0,255}

\newcommand{\needtocheck}[1]{{#1}}

\newcommand{\majorrev}[1]{#1}

\newcommand{\minorrev}[1]{#1} 

\newcommand{\ipstart}[1]{\vspace{1mm} \noindent{\textbf{\textit{#1.}}}}

\newcommand{\circledigit}[1]{\textbf{\normalsize{\textsf{\textcircled{\footnotesize{#1}}}}}}

\definecolor{tableheader}{HTML}{EFEFEF}
\definecolor{tablegrayline}{HTML}{d0d0d0}

\definecolor{darkgray}{HTML}{555555}

\newcommand{\labelphantom}[1]{%
  \parbox{0pt}{\phantomsubcaption\label{#1}}%
}

\newcommand{\symbolbot}{\raisebox{-2.3pt}{\includegraphics[width=10pt]{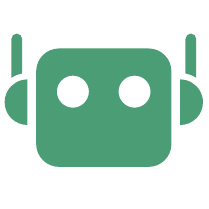}} }

\newcommand{\symboluser}{\raisebox{-2.1pt}{\includegraphics[width=9.5pt]{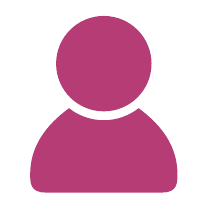}} }

\newcommand{\siglegend}{\sffamily\footnotesize{\textbf{***}$p$<.001; \textbf{**}$p$<.01; \textbf{*}$p$<.05}}

\makeatletter
\if@todonotes@disabled

\else

\fi
\makeatother

\usepackage{caption}
\usepackage{subcaption}
\usepackage{tabularx}
\usepackage{multirow}
\usepackage{xcolor}
    
\usepackage{newfloat}
\DeclareFloatingEnvironment[fileext=fml,placement={H},name=Fragment]{fragment}
\AtBeginDocument{%
  \providecommand\BibTeX{{%
    \normalfont B\kern-0.5em{\scshape i\kern-0.25em b}\kern-0.8em\TeX}}}

\setcopyright{acmcopyright}
\copyrightyear{2023}
\acmYear{2023}
\acmDOI{XXXXXXX.XXXXXXX}

\acmConference[CSCW 24']{27th ACM Conference on Computer-Supported Cooperative Work and Social Computing}{November 9-13,
  2024}{San José, Costa Rica
}
\acmBooktitle{(CSCW ’24), San José, Costa Rica} 
\acmPrice{15.00}
\acmISBN{978-1-4503-XXXX-X/18/06}

\begin{document}

\title{Leveraging Large Language Models to Power Chatbots for Collecting User Self-Reported Data}

 \author{Jing Wei}
 \authornote{Jing Wei conducted this work as a research intern at NAVER AI Lab.}
 \email{jwwei2@student.unimelb.edu.au}
 \affiliation{%
 \institution{University of Melbourne}
 \city{Parkville}
 \state{VIC}
 \country{Australia}
 }

 \author{Sungdong Kim}
 \email{sungdong.kim@navercorp.com}
 \affiliation{%
 \institution{NAVER AI Lab}
 \city{Seongnam}
 \state{Gyeonggi}
 \country{Republic of Korea}
 }

 \author{Hyunhoon Jung}
 \email{hyunhoon.j@navercorp.com}
 \affiliation{%
 \institution{NAVER CLOUD}
 \city{Seongnam}
 \state{Gyeonggi}
 \country{Republic of Korea}
 }

 \author{Young-Ho Kim}
 \email{yghokim@younghokim.net}
 \orcid{0000-0002-8522-8607}
 \affiliation{%
 \institution{NAVER AI Lab}
 \city{Seongnam}
 \state{Gyeonggi}
 \country{Republic of Korea}
 }

\renewcommand{\shortauthors}{Jing Wei, Sungdong Kim, Hyunhoon Jung, and Young-Ho Kim}

\begin{abstract}

Large language models (LLMs) provide a new way to build chatbots by accepting  natural language prompts. Yet, it is unclear how to design prompts to power chatbots to carry on naturalistic conversations while pursuing a given goal such as collecting self-report data from users. We explore what design factors of prompts can help steer chatbots to talk naturally and collect data reliably. To this aim, we formulated four prompt designs with different structures and personas. Through an online study ($N$ = 48) where participants conversed with chatbots driven by different designs of prompts, we assessed how \textit{prompt designs} and \textit{conversation topics} affected the conversation flows and users' perceptions of chatbots. Our chatbots covered 79\% of the desired information slots during conversations, and the designs of prompts and topics significantly influenced the conversation flows and the data collection performance. We discuss the opportunities and challenges of building chatbots with LLMs.
\end{abstract}




\begin{teaserfigure}
  \centering
  \includegraphics[width=0.9\textwidth]{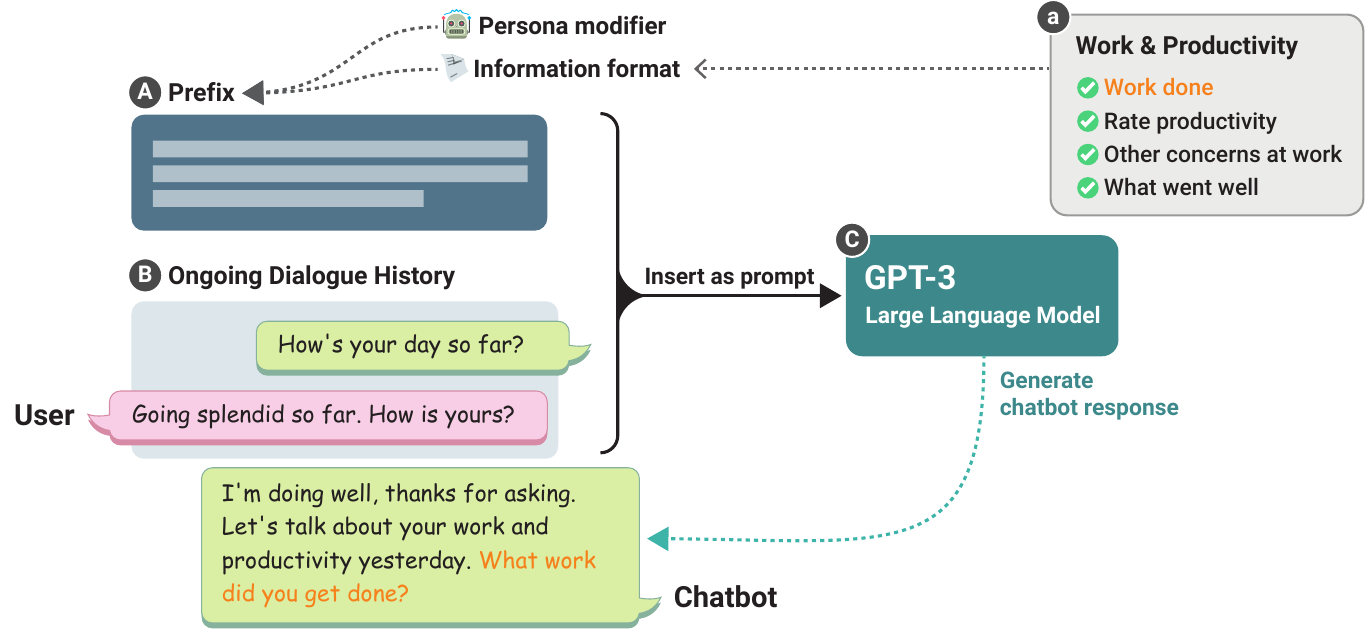}
  \caption{An overview of our chatbot running on a large language model through zero-shot response generation, with only a prefix \circledigit{A} consisting of persona modifier and information format and the ongoing dialogue history \circledigit{B}. The example conversation is carried on about \textit{work} \circledigit{a}. GPT-3 was used for an underlying large language model \circledigit{C}. 
  \needtocheck{} }
  \label{fig:teaser}
  \Description{Teaser image of ...}
\end{teaserfigure}

\maketitle

\section{Introduction}

Conversational Agents (CAs) or chatbots are gaining a wide popularity. Applications such as Apple Siri~\cite{AppleSiriHealth}, Amazon Alexa~\cite{AmazonAlexa}, and Google Assistant~\cite{GoogleAssistant} are becoming ubiquitous. Suggested by its name, CAs interact with people in natural languages, thereby providing customer services~\cite{cui2017superagent} and companionship~\cite{ta2020user}. Compared to GUIs, CAs have shallower learning curves and may even form a relationship with people through conversations~\cite{dingler2021use}. The digital health domain particularly benefits from the use of CAs: chatbots can be deployed to collect self-reports and provide personalized coaching to different individuals~\cite{mitchell2022examining}. Studies have found that people are willing to engage with chatbots and provide valuable information, such as self-reports~\cite{wei2021understanding} and survey responses~\cite{celino2020submitting}, to chatbots~\cite{xiao2020if, xiao2020tell}.

Many digital health applications that are designed to promote behavioral changes and health interventions require people's long-term adherence. Chatbots' ability to ``converse'' naturally, as distinguished from the GUI-based systems~\cite{luger2016like}, has the potential to help people develop long-term adoptions for health monitoring~\cite{dingler2021use, fitzpatrick2017delivering, mitchell2022examining}. However, existing commercial chatbot frameworks, such as Dialogflow\cite{DialogFlowDev} and Amazon Alexa~\cite{AlexaDev}, predominantly only support building rule-based and scripted chatbots~\cite{wei2021understanding, luo2020tandemtrack}. Lacking flexible flows, these chatbots usually appear robotic and unnatural~\cite{mctear2018conversational}. \majorrev{Particularly, using rule-based chatbots to collect user reports may} cause boredom in long-term deployments\majorrev{~\cite{tian2021let}}. On the other hand, implementing chatbots that can have more diverse and dynamic conversations requires large and specific domain datasets~\cite{mitchell2022examining}. For example, an open-domain chatbot Meena was trained on 341 GB of dialogue sessions~\cite{adiwardana2020towards}. Since creating such large datasets is costly, the datasets are often proprietary and inaccessible publicly. Furthermore, most demonstrations of open-ended chatbots focus on performing free-form conversations in general topics and do not support end-user customizations. Little research has been done to explore low-effort bootstrapping ways to build chatbots that can effectively perform pre-defined tasks, such as inquiring people about their health information and carrying on naturalistic conversations at the same time.

Recent Large Language Models (LLMs; \eg, GPT-3~\cite{brown2020language}, PaLM~\cite{Chowdhery2022PALM}, OPT~\cite{Zhang2022OPT}, HyperCLOVA~\cite{Kim2021HyperCLOVA}), with billions of parameters pre-trained on a large amount of language corpora, provide new \majorrev{opportunities for conversational agents. The recently released ChatGPT has attracted \minorrev{over} 1 million users within five days~\cite{altman_2022}. ChatGPT has exceeded many people's expectations by showing its vast amount of knowledge and the ability to converse in natural languages~\cite{review_2022}. \minorrev{GPT-3.5 and 4}, the backend models of ChatGPT, can be further fine-tuned and prompted to build specified chatbots that perform tasks such as acting as a virtual coach to effectively inquire people for certain information. By designing specific prompts, we can make GPT-3, or LLMs in general, produce human-like conversation} responses accordingly without any training data, thereby functioning as a chatbot. Compared to other frameworks, LLMs show great potential in scaffolding chatbots that are sensible of contexts and even respond to off-topic user messages~\cite{volum2022craft}. Further, \majorrev{LLMs may change the way of people cooperate since they operate on natural language inputs. Previously, building chatbots may be limited to people with technical background. But LLMs enable people, such as medical practitioners, to} have the opportunity to personalize or even build their own chatbots~\cite{dingler2021use}.

Despite these potentials, it is yet fully understood how LLMs \textit{read} the prompt and \textit{use} pretrained knowledge~\cite{brown2020language, Liu2021RetrievalGPT3}, the development of prompts is usually conducted through iterative trial and error~\cite{liu2022design}. While the HCI and CSCW community have actively explored the use of LLMs in various domains (\eg,~\cite{chung2022talebrush, lee2022coauthor, wu2022ai}), research that leverages LLMs for powering chatbots, particularly task-oriented ones~\cite{mehri2022lad, bae2022building, volum2022craft}, is still sparse. Due to the inherent characteristics of LLMs, LLM-driven chatbots may be error-prone~\cite{korngiebel2021considering} or digress from their tasks~\cite{volum2022craft}. Designing robust prompts is crucial for ``restricting'' chatbots to conduct desired tasks.

In this study, we investigate how LLMs can power chatbots to collect user self-reports while carrying on naturalistic conversations. Towards this aim, we built a set of chatbots (\autoref{fig:teaser}) that run on GPT-3~\cite{brown2020language} and converse to collect self-report data in four health-related topics---sleep, food intake, work and productivity, and exercise. We chose GPT-3 as an underlying LLM because it is one of the mainstream LLMs that are publicly available via commercial APIs. We formulate the model prompt to include the information slots (\ie, information properties of a topic) that we intend the chatbot to collect and the job identities (\eg, sleep expert for the topic sleep) to help drive the conversations. We investigate how two design factors in prompts---information specification format and personality modifier---impact the \majorrev{\textit{slot filling ability}} and the \majorrev{\textit{conversation style}} of chatbots. In total, we created 16 chatbots (4~topics $\times$ 2~formats $\times$ 2~personality modifiers) with different prompts.

\majorrev{To the best of our knowledge, our work is the first to explore the usefulness of LLMs in powering chatbots for collecting self-report data. We believe that well-designed prompts can effectively drive chatbots to perform specified tasks~\cite{wang2023enabling}. In the context of data collection through conversations, we evaluate chatbots from two perspectives: (1) slot filling performance, and (2) conversational styles. W}e conducted an online study ($N$ = 48) with our chatbots on a web interface. All participants talked to chatbots of the four topics but each of them experienced chatbots run on the same prompt design. 
To the best of our knowledge, our work is the first to explore the usability of LLMs for building chatbots for collecting self-report data. We found that our zero-shot prompts, without either example dialogues or fine-tuning, covered 79\% of the desired information slots among all dialogues. Through conversation analysis, we found that the information specification format as well as the use of personality modifier can impact the chatbots' slot-filling ability and conversation styles. Also, the chatbots generally reacted to participants' self-reported answers in an empathetic way, appreciating their accomplishments as well as sympathizing with participants for the negative outcomes. Consequently, some participants perceived these chatbots to be understanding and take into account their messages when responding, and others indicated that they were surprised to find the chatbots' responses were accurate and detailed. 

\vspace{5pt}\noindent{}The contributions from this work are threefold:
\begin{enumerate}[leftmargin=*, itemsep=4pt, topsep=3pt]
    \item Empirical results from a \minorrev{between-subject} online study ($N=48$), demonstrating the \textit{feasibility} of chatbots powered by LLMs in not only carrying on conversations to collect specified information but also exhibiting abilities in maintaining context, state-tracking, and providing off-topic suggestions.
    \item Examination of how different prompt designs and other factors impact the chatbots' behaviors, providing insights for future researchers to easily scaffold chatbots \minorrev{through zero-shot prompting} for data collection with LLMs.
    \item Implications on how future LLMs-driven chatbot platforms can improve the conversation quality, drawing on the analysis of the dialogue errors.
    \end{enumerate}
\section{Related Work}
In this section, we cover related work in the areas of (1) self-report data collection through chatbots, (2) design considerations for chatbots, (3) chatbot platforms, and (4) designing LLM prompts for chatbots.

\subsection{Self-Report Data Collection through Chatbots}
Personal informatics systems have commonly incorporated data collection techniques to track personal health and activity~\cite{Li10StageBasedModel, Choe2017SemiAutomatedTracking}.
While various physiological or physical activity data---such as step count, heart rate, and sleep duration---can be captured automatically by sensors and wearable devices~\cite{laborde2020older}, various types of personal data still demand \textit{self-reporting} by the person who self-tracks~\cite{Choe2017SemiAutomatedTracking}. For example, food intake (\eg, ~\cite{cordeiro2015barriers}) or work tasks (\eg,~\cite{Kim2019UnderstandingProductivity}) are not reliably captured by sensors and thus require manual inputs. In addition, reflective questions (\eg, Why did you eat this food?~\cite{luo2021foodscrap}) and subjective measurements (\eg, Sleep quality) inherently require to be captured manually. 
A majority of digital self-tracking tools that involved manual data capture inherited the traditional concepts of self-monitoring or journaling and provide form-based GUIs such as a list of checkboxes and text fields~\cite{Kim2017OmniTrack, Jeon2016MsThesis}. However, repeated manual input on a computer or smartphone screen is burdensome and may gradually disengage people from tracking~\cite{choe2014understanding, Choe2017SemiAutomatedTracking}. 
As an input modality to lower the capture burden and enhance the richness of the captured information, natural language has recently gained interest~\cite{luo2021foodscrap, Silva2021Food}. Prior research found that when people are allowed to insert data in free-form natural language, they tend to provide detailed answers with surrounding contexts~\cite{luo2021foodscrap, Kim2022MyMove}. 
Going further, conversational interaction, where a system and a user communicate in natural languages, has become one emerging interface for collecting self-reports.

Chatbots are considered easier to use and more accessible than GUIs as they minimize the use of graphical widgets employ the intuitive conversational interaction. Regarding data collection, a plethora of research has explored the use of chatbots in place of traditional form-based surveys (\eg,~\cite{xiao2020tell, bemmann2021chatbots, celino2020submitting, kim2019comparing}). For example, studies with surveys with close-ended questions found that chatbots can collect the same quality, if not higher, user responses as GUIs~\cite{celino2020submitting, kim2019comparing}. \citet{xiao2020tell} built a chatbot to conduct interviews with open-ended questions. Compared to the traditional web survey, their participants showed higher engagement and provided higher-quality responses when talking to the chatbot. Further, incorporating more humanized traits, such as casual conversation styles~\cite{kim2019comparing}, self-introduction, and echoing~\cite{rhim2022application}, led to not only a higher level of user engagement and satisfaction but also more self-disclosure in responses.
With more focus on self-reported data, prior studies leveraged chatbots to collect self-reports such as emotion (\eg,~\cite{bemmann2021chatbots}), pain level (\eg,~\cite{wrzus2015lab}), and food intake (\eg,~\cite{mitchell2022examining}). For example, \citet{bemmann2021chatbots} combined a chatbot with the experience sampling method (ESM, \cite{larson2014experience}) and found that personalized chatbots have the potential to collect data on sensitive or personal topics. 
\citet{mitchell2022examining} compared fully-scripted, rule-based, and retrieval-based chatbots for collecting food nutrition. They found the better fulfillment of data collection is not necessarily associated with the higher perceived quality of the chatbot as a diet coach, suggesting the importance of conversational content in user experience.

This work expands the line of research on chatbots that collect self-reports. In contrast to prior studies that involved predefined conversation logic or retrieval model training on domain-specific datasets, 
we explore the potential of LLMs in bootstrapping chatbots that can collect self-reports through conversations on four health topics---sleep, food, work, and exercise.  

\subsection{Design Considerations for Chatbots}
Prior works in HCI explored user behaviors with chatbots and proposed suggestions to improve user experience with them. For example, ~\citet{luger2016like} found that people restricted their language uses when interacting with CAs. ~\citet{jain2018evaluating} revealed that many first-time chatbot users had disappointment and frustration with the selected chatbots: most chatbots lacked the ability to fully comprehend user messages or intentions. Since conversation breakdowns are still common~\cite{ashktorab2019resilient}, several studies have explored repair strategies, such as apologies, compensation, and providing option~\cite{lee2010gracefully}. ~\citet{ashktorab2019resilient} also evaluated other strategies, such as confirmation, repeat, keywords highlight \& explanation, and recommended that chatbots should acknowledge misunderstanding in simple terms, explain model limitation in natural ways, and adapt individualized strategies. Although existing chatbot frameworks also have error recovery features, their features are not only limited but often cannot allow quick repairs~\cite{wei2022could, myers2018patterns, mctear2018conversational}. 

Another key to improving the user experience is to make chatbots more playful and human-like~\cite{liao2018all}. The level of empathy~\cite{celino2020submitting, rashkin2018towards} and the repetitive rate~\cite{see2019makes} are two commonly used metrics of human-likeness. For example, the playful interactions (\eg, telling jokes) or humorous responses enabled many people to start using CAs~\cite{luger2016like} and it is crucial for chatbots to support sustainable playfulness~\cite{serenko2008model}. Also, human-like features and fun personalities are found to make chatbots more enjoyable to interact~\cite{jain2018evaluating}. Even for work-related chatbots, some people still preferred chatbots that were human-like~\cite{liao2016can}, and ~\citet{liao2018all} envisioned that a reusable conversational module including common chit-chats and social attributes could be developed. In other words, future chatbot platforms should allow developers to easily build personalized chatbots with different personalities~\cite{volkel2020developing} and conversation styles~\cite{braun2019your, krenn2014effects}. Lastly, developers should aim to improve chatbots' ability to maintain contexts to support smoother and natural conversations~\cite{ashktorab2019resilient}. In this work, we investigate whether LLMs can steer chatbots that have social attributes and can resolve conversation breakdowns.

\subsection{Chatbot Platforms}
Building chatbots is challenging and time-consuming, and many design suggestions discussed above are difficult to implement. Many open-domain chatbots that engage and entertain people socially are \minorrev{predominantly} dependent on large datasets~\cite{zhou2020design, adiwardana2020towards}. In the HCI and CSCW community, rule-based dialogue systems are widely used. \citet{celino2020submitting} built their survey chatbot with pre-defined conversation flows as they intended to avoid disappointments caused by the chatbot's inability to understand certain utterances~\cite{luger2016like}. Although rule-based chatbots are unlikely to cause breakdowns, the resulted rigid conversations can make people lose interest in the long term. On the other hand, ~\citet{xiao2020tell} built their survey chatbot using Juji~\cite{Jujiio}, which automatically equip chatbots with rich existing conversational skills. Using the Juji GUI to add questions is relatively simple, but it is unclear whether developers can modify the chatbot's expressed personality. Lastly, other commercial chatbot frameworks, such as Dialogflow~\cite{GoogleDialogFlow} and IBM Watson~\cite{IBMWatson}, also allow developers to build rule-based chatbots with GUIs~\cite{thorat2020review}. However, creating more dynamic conversations usually requires programming skills. Even for professional developers, it is challenging to create well-designed conversational flows and pre-define user intents and chatbot messages~\cite{mctear2018conversational}. Using LLMs to power chatbots is a new way to build chatbots~\cite{volum2022craft, brown2020language}. LLMs accept natural language prompts so that people without any knowledge of programming but are interested in building chatbots for data collection can create prompts~\cite{Kim2021HyperCLOVA}. Customizing prompts in natural languages essentially hands off the control to each individual who builds chatbots. As such, it becomes more straightforward to scaffold personalized chatbots (\eg, assigning a preferred personality) by revising prompts accordingly. Nevertheless, it is unclear how to design prompts for LLMs to steer chatbots that can effectively ask questions around desired information and have different conversational styles.

\subsection{Designing LLM Prompts for Conversations}

Prompts are natural language texts to LLMs to produce desired outputs. With proper prompt inputs, GPT-3 can be used to translate texts, answer questions, write essays, and generate dialogues without any fine-tuning~\cite{brown2020language}. While the mechanism enabling such few-shot abilities behind LLMs is still veiled~\cite{brown2020language, min2022rethinking}, some prompting techniques are found to improve the model performance. One technique that surprisingly improves the generation quality is by conditioning the prompt with an identity. For example, by inserting the statement ``\texttt{You are an expert Python programmer}'' into prompts, models can generate higher quality codes~\cite{austin2021program}, and similarly, ``\texttt{I'm a math tutor}'' is suggested to improve models' performance in solving math problems~\cite{austin_2022}. In HCI, researchers are also interested in investigating the design guidelines for prompts. ~\citet{liu2022design} explored different design factors, such as phrasing, styles, subjects, and random seeds of prompts for text-to-image generative models. They found that the subject and style keywords are more important than the connecting words and the initializing seeds can significantly impact the quality of generations. Similarly, ~\citet{lee2022coauthor} discovered that the randomness impacted writing collaboration and new identity generations. In terms of dialogues, some studies designed prompts for LLMs to generate single-turn responses~\cite{Wang2021TheraphyChatbot} and examined whether these artificial responses are comparable to human responses~\cite{tack2022ai}.


\majorrev{\minorrev{More} recently, OpenAI released ChatGPT in November 2022 and its API subsequently in March 2023. ChatGPT is optimized to provide a conversational interface for users to interact with the LLM. The ease-of-use of ChatGPT attracts the attention of billions of users~\cite{chatgpt-aibusiness}. Practitioners have actively proposed open-source tools based on the ChatGPT API, such as LangChain~\cite{langchain} and AutoGPT~\cite{autogpt}, which support building conversational agents~\cite{vaghefi2023chatclimate}. Researchers have also explored the usefulness of LLM-driven chatbots for \minorrev{various types of end-users (\eg, programmers~\cite{ross2023programmer}, socially-isolated people~\cite{Jo2023Benefits}, and children~\cite{seo2023chacha}) as well as tasks (\eg, UI task~\cite{wang2023enabling}, medical information search~\cite{lee2023benefits})}. Without fine-tuning, prompts are the key to drive chatbots converse effectively. However, despite providing used prompts, prior work did not investigate how to optimize prompts for domain-specific chatbots. Further, compared to previous task-oriented chatbots that mostly answer user questions, our study aims to build chatbots that can proactively ask users pre-defined questions. The challenge of our study is that we need to design prompts to power chatbots that (1) ask specific questions with domain knowledge and (2) can converge conversations on their own.}

\majorrev{In summary, a plethora of prior works have leveraged LLMs to support conversations. However, few works has explored prompting chatbots that can \textit{lead} conversations and perform question-asking tasks.} In this study, we set out to address this gap and provide empirical insights into prompt designs for chatbots with different ``job identities'' that are dedicated to collecting self-reports. Further, we investigate whether LLM-driven chatbots can exhibit important features such social attributes~\cite{liao2018all}, employ empathy~\cite{rashkin2018towards}, as well as have the ability to handle breakdowns~\cite{ashktorab2019resilient}.
\section{Method}
To examine how different prompt design factors impact LLM-driven chatbots' performance, we implemented a web-based chatbot interface and conducted an online user study to collect dialogue data from people. We describe our prompt designs and the experimental design of the user study.

\subsection{The LLM-driven Chatbot Framework}
LLMs power chatbots by generating the next response based on the input prompt. \autoref{fig:teaser} describes the mechanism of our chatbot running on an LLM. At each turn, the back-end system combines a prefix (\circledigit{A} in \autoref{fig:teaser}) and the history of the ongoing conversation (\circledigit{B} in \autoref{fig:teaser}) into a single prompt and feeds it into the LLM (\circledigit{C} in \autoref{fig:teaser}). Then the LLM generates the following agent's response. The prefix, treated as an instruction for an LLM, consists of the description of persona and the specification of the desired information slots that the chatbot should capture.

\subsubsection{Prompt Designs for Chatbots}

\begin{table*}[t]
    \small\sffamily
			\def\arraystretch{1.5}
		    \setlength{\tabcolsep}{0.3em}
		    \centering
\caption{The specified information slots and job identity for each conversation topic.}
\begin{tabular}{|p{0.14\textwidth}!{\color{lightgray}\vrule}p{0.13\textwidth}!{\color{lightgray}\vrule}p{0.69\textwidth}|}
\hline
\rowcolor{tableheader}
\textbf{Topic} & \textbf{Job Identity} & \textbf{Information Slots} \\ \hline
\textbf{Sleep} & Sleep Expert & \parbox[t]{0.65\textwidth}{\slotbox{(1) Time to bed} \slotbox{(2) Sleep latency} \slotbox{(3) Wake up at night}\\\slotbox{(4) Wake up time} \slotbox{(5) Sleep quality rating (1--10)}} \\\arrayrulecolor{tablegrayline}\hline
\textbf{Food Intake} & Dietitian & \slotbox{(1) Breakfast} \slotbox{(2) Lunch} \slotbox{(3) Dinner} \slotbox{(4) Snacks} \slotbox{(5) Feelings after eating} \\\arrayrulecolor{tablegrayline}\hline
\parbox[t]{0.15\textwidth}{\textbf{Work and\\Productivity}} & Life Coach & \parbox[t]{0.69\textwidth}{\slotbox{(1) Work done} \slotbox{(2) Productivity rating (1--10)} \slotbox{(3) Other concerns at work} \slotbox{(4) What went well}} \\\arrayrulecolor{tablegrayline}\hline
\textbf{Exercise} & Fitness Coach & \parbox[t]{0.69\textwidth}{\slotbox{(1) What workout} \slotbox{(2) Workout duration}\\ \slotbox{(3) Feeling after (skipping) workout} \slotbox{(4) Fitness concerns}} \\\arrayrulecolor{black}\hline

\end{tabular}
\label{tab:slots}
\end{table*}

In this study, we envisioned a self-monitoring scenario where chatbots proactively interact with people and inquire them about their retrospective health-related behaviors of the previous day through natural conversations. Towards this goal, we first chose four health-related topics---\textbf{Sleep}~\cite{min2014toss}, \textbf{Food} intake~\cite{luo2021foodscrap, Epstein2016Taking5}, \textbf{Work} and productivity~\cite{li2011understanding, Kim2019UnderstandingProductivity, Kim2016TimeAware}, and \textbf{Exercise}~\cite{luo2020tandemtrack}---that retain interests not only by the research community and are also personal use in daily life~\cite{williams2006simply}. Then, we defined the behavioral data (slots) that we intended to have chatbots to collect. \autoref{tab:slots} summarizes the information slots for each topic, which the chatbots should capture in each conversation session. Another aspect that we intended to explore was whether we could enable LLMs to steer chatbots to exhibit different conversation styles as different individual's experience with chatbots can be improved with personalizations~\cite{braun2019your}.

Currently, there is a lack of prompt designs \majorrev{in general, let alone for chatbots.  As suggested by earlier work, even NLP experts adopt trial and error and iterative experimentation when designing prompts before applying prompts in large datasets~\cite{zamfirescu2023johnny}. The designs of prompts largely depend on the intended task. The goal of this study is to understand how different prompts impact the data collection performance of chatbots. Prior to deploying GPT-3-powered chatbots to participants, we trialed several iterations of prompts to find satisfactory chatbot prompts among our research team. In our ``trial and error'' process, we judged prompts based on whether the resulted chatbots can fulfill the assigned tasks and not go off-topic.} 

Initially, we started with OpenAI's example prompt, which defines an ``AI assistant'' with several characteristics(\eg, ``\texttt{The following is a conversation with an AI assistant. The assistant is helpful, creative, clever, and very friendly.}''). We modified the example prompt by including questions of interests (\eg, ``\texttt{How's your sleep last night?}'') as well as by adding the list of information slots (\eg, ``...\texttt{The assistant asks questions about human's sleep time, sleep duration, and sleep quality.}''). However, we found that these modifications tended to make conversations digress and the resulted chatbots converse more similar to generic customer service chatbots.

Findings from research~\cite{Kojima2022ZeroShotReasoners, wei2022chain, austin2021program} and anecdotal evidences from social medias~\cite{austin2021program} suggest that specifying identities (\eg, ``Python programmer'') can improve GPT-3 model performance. Inspired by this, we substituted the ``AI assistant'' with more specific job identities (\eg, ``fitness coach'') and found that the resulted chatbots appeared to have more domain knowledge and are less likely to digress in our tests. As such, we picked different job identities for each topic---\textit{sleep expert} (Sleep), \textit{dietitian} (Food), \textit{life coach} (Work), and \textit{fitness coach} (Exercise).

\begin{figure*}
     \centering
     \includegraphics[width=\textwidth]{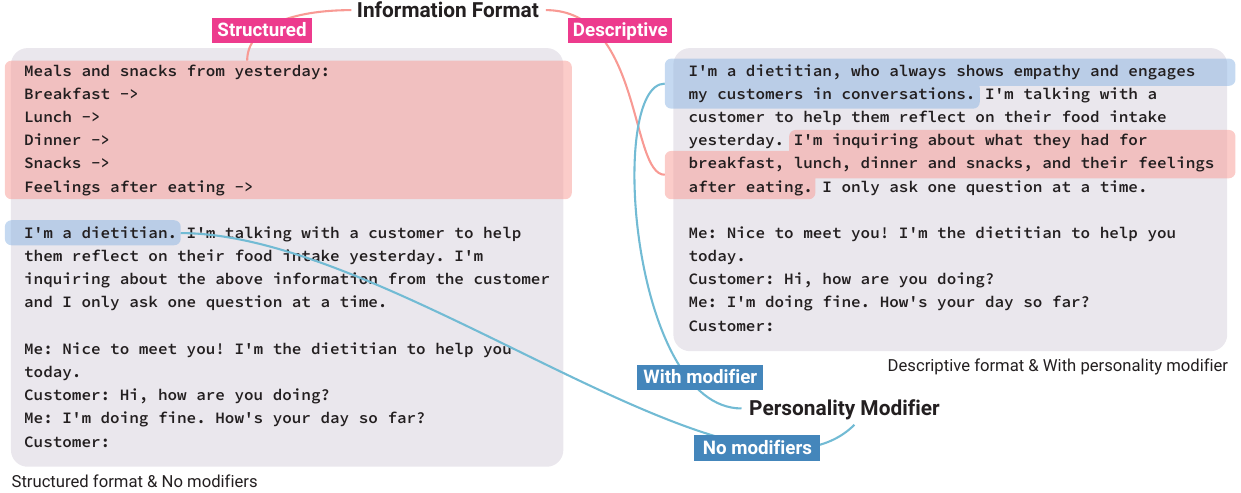}
     \caption{Prompt design combining two factors, information format and personality modifier, in the Food intake topic.}
     \label{fig:chatbot_prompt}
\end{figure*}

With the above modifications, we designed prompts by first describing a chatbot with a specific identity and then including a list of slots (\eg, ``\texttt{I'm inquiring about what they had for breakfast, lunch, dinner, and snacks...}''; see \autoref{fig:chatbot_prompt}, right). Through multiple trials, this design worked better than generic one despite still had occasional digressions. Hence, we experimented with another prompt format inspired by the state-tracking technique in task-oriented dialogues~\cite{Lee2021Dialogue}. Instead of being described literally, the slots are structured into a form (\eg, ``\texttt{Meals and snacks from yesterday: Breakfast -> [placeholder] Lunch -> [placeholder] ...}''; we used an empty string as a placeholder; see \autoref{fig:chatbot_prompt}, left). Both designs performed similarly in our limited internal testings, hence we aimed to investigate the performance of two formats in the user study.

In terms of manipulating conversational styles, we introduced the use of a modifier in prompts--``\texttt{who always shows empathy and engages my customer in conversations},'' to the prompt (See \autoref{fig:chatbot_prompt}, right). We hypothesize that with this modifier, the chatbot is more likely to express empathy in conversations and have a higher level of interactivity---\ie, use more emphatic expressions and be more responsive to user responses. Conversely, without the modifier (See \autoref{fig:chatbot_prompt}, left), we expect the chatbot to be more neutral, formally exchanging messages with users and appear less empathetic. Lastly, during our trials, we found that GPT-3 had the tendency to ask multiple questions in one turn. To restrict this behavior, we added ``\texttt{I only ask one question at a time.}'' to the prompt.

\subsubsection{Model and Parameters}
To power chatbots with above prompts, we used \texttt{davinci-text-002}, the largest and most capable model of GPT-3 as of June 2022, publicly accessible via OpenAI's API~\cite{openai}.
This model accepts 4,000 byte-pair encoding tokens at maximum in a prompt per request. Our prompt templates in the initial state were encoded into around only 120 tokens (3\% of the limit) and allowed sufficient room for the appended conversation history.
For all chatbots, we uniformly applied the same generative parameters: temperature as 0.9, the presence penalty as 0.6, and the frequency penalty as 0.5. We kept the temperature and the presence penalty unchanged based on OpenAI's suggestions and increased the frequency penalty to reduce the re-use of words.

\subsubsection{The Web Chat Interface}
We implemented a web interface to host our LLM-driven chatbots, following a typical chat interface design (See Appendix \ref{appendix:interface}). The webpage was written in TypeScript~\cite{TypeScript} on React~\cite{React} and runs on the Node.js~\cite{nodejs} server. The server communicates with GPT-3 leveraging OpenAI's API~\cite{openai}. To simplify the conversation flow, we disabled people to submit multiple utterances in a row. Correspondingly, the chatbots also delivered one utterance at a time. When a user submitted an utterance, the server appended the current dialog history at the end of the prompt template and fed it to GPT-3 to generate the following response. 

\subsection{Online Study}

\subsubsection{Experimental Conditions}
Combining the two design factors, we created four designs of prompts: SP (\textbf{S}tructured format with \textbf{P}ersonality modifier), SN (\textbf{S}tructured format with \textbf{N}o personality modifier), DP (\textbf{D}escriptive format with \textbf{P}ersonality modifier), and DN (\textbf{D}escriptive format with \textbf{N}o personality modifier). Each participant was assigned to one prompt design and engaged in conversations of all four topics. (Refer to the supplementary material for all 16 variations of GPT-3 prompts created for combinations on topic and condition.) To mitigate the ordering effect among topics, half of the participants conversed in the order of Work--Food--Exercise--Sleep, and the other half in the order of Exercise--Sleep--Work--Food. Additionally, for each topic, we requested participants to engage with the chatbot twice: one in the \textbf{Positive} path (\eg, report high-quality sleep) and one in the \textbf{Negative} path (\eg, report poor sleep). \majorrev{The two paths served as a general guideline for participants to exercise their imaginations in conducting \textit{open-ended} conversations with the chatbots. Participants could make up their own ``stories'' and we aimed to investigate whether and how GPT-3-powered chatbots can handle all sorts of user responses on their \textit{own}. }
Refer to Appendix \ref{appendix:paths} for an exhaustive list of paths and hints by topic provided by us to guide participants to compose their answers for each path accordingly.

\subsubsection{Web Chat Session}
After signing an electronic consent form on the study website, participants went through eight conversations ($4~topics * 2~paths$). On the web chat interface (See Appendix \ref{appendix:interface}), we put guidelines including the instructions and the conversation path that participants should follow (See Blue text in Appendix \ref{appendix:interface}, right). Since we did not incorporate ending detection algorithms, we asked participants to click the `Next' button to proceed to the next conversation when they thought the conversation was naturally over or the chatbot kept sending repetitive messages. The completed dialogues were stored in our server.

\subsubsection{Exit Survey}
After completing eight conversations, the web page automatically redirected participants to an online survey. The survey consisted of three 5-point Likert scale questions and one open-ended feedback textfield. The Likert scale questions were: (1) ``\textit{Do you think the chatbot understands your answers?}'' (2) ``\textit{Do you think the chatbot takes into account of your answers when responding?}'' and (3) ``\textit{Do you think the chatbot talks more like a human who shows more empathy or more like a robot who behaves mechanically?}'' The open-ended feedback question stated, ``\textit{If you have any other comments or thoughts about the chatbot (\eg, things that you've liked or disliked), please share with us.}'' 
The first two questions can measure whether participants think the chatbot acknowledges their answers and respond accordingly and the third question is an overall measure of whether the chatbot is perceived as being empathetic. With participants' subjective evaluations, we hope to see whether the personality modifier can impact the chatbot's way of talking.

\subsection{Participants}
We recruited participants by word-of-mouth and posting advertisements at a large tech company, social media, and online forums in local universities. We sent the link to our study website to 83 people who filled out a screener and met our inclusion criteria: (1) aged 19 or older; (2) fluent English speaker; and (3) have the experience in talking to chatbots of any kind. 54 people completed the online study session and submitted an exit survey. The entire study lasted less than 20 minutes and all participants received e-gift cards (equivalent to \$5 USD) after they completed the study. 

We excluded six people's data from analysis; one made significant amount of grammatical errors and the rest completed less than half dialogues. \autoref{tab:demographics} summarizes the demographic of the final 48 participants (aged 19 to 56, 18 females). Fourteen out of 48 (29\%) participants were native/bilingual and 22 out of 48 (46\%) participants had never heard of or used LLMs. Each prompt design condition included 12 participants.

\begin{table*}[t]
\small\sffamily
			\def\arraystretch{1.2}
		    \setlength{\tabcolsep}{0.6em}
		    \centering
\caption{Participant demographics by experimental condition.}
{%
\begin{tabular}{lr*{4}{m{0.08\linewidth}}}
\hline
\rowcolor{tableheader}
& & \multicolumn{1}{c}{\textbf{SP}} & \multicolumn{1}{c}{\textbf{SN}} & \multicolumn{1}{c}{\textbf{DP}} & \multicolumn{1}{c}{\textbf{DN}} \\ \hline
\textbf{Age} & Mean (min--max)                                &    \multicolumn{1}{c}{31.5 (21--56)}                &  \multicolumn{1}{c}{28.0 (21--33)}             &  \multicolumn{1}{c}{27.5 (20--40)}                  &  \multicolumn{1}{c}{30.4 (19--42)} \\\arrayrulecolor{tablegrayline}\hline
\multirow{2}{*}{\textbf{Gender}} & Male                                           &         \multicolumn{1}{c}{7}                &    \multicolumn{1}{c}{8}                 & \multicolumn{1}{c}{8}                     &  \multicolumn{1}{c}{7} \\
& Female                                       &       \multicolumn{1}{c}{5}                 &   \multicolumn{1}{c}{4}              &     \multicolumn{1}{c}{4}                   &  \multicolumn{1}{c}{5} \\ \hline
\multirow{2}{*}{\begin{tabular}[c]{@{}l@{}}\textbf{English}\\\textbf{Proficiency}\end{tabular}} & Native/Bilingual               &      \multicolumn{1}{c}{4}     &     \multicolumn{1}{c}{2}    &      \multicolumn{1}{c}{4}  &   \multicolumn{1}{c}{4} \\
& Proficient                                             &      \multicolumn{1}{c}{8}     &     \multicolumn{1}{c}{10}    &      \multicolumn{1}{c}{8}  &   \multicolumn{1}{c}{8}   \\ \hline
\multirow{4}{*}{\textbf{Education}} & High school                                        &      \multicolumn{1}{c}{1}     &     \multicolumn{1}{c}{2}    &      \multicolumn{1}{c}{1}  &   \multicolumn{1}{c}{1}  \\
& Bachelor                                           &      \multicolumn{1}{c}{5}     &     \multicolumn{1}{c}{4}    &      \multicolumn{1}{c}{3}  &   \multicolumn{1}{c}{3}  \\

& Master                                             &      \multicolumn{1}{c}{4}     &     \multicolumn{1}{c}{5}    &      \multicolumn{1}{c}{7}  &   \multicolumn{1}{c}{7} \\

& Doctor                                             &      \multicolumn{1}{c}{2}     &     \multicolumn{1}{c}{1}    &      \multicolumn{1}{c}{1}  &   \multicolumn{1}{c}{1}  \\ \hline
\multirow{4}{*}{\begin{tabular}[c]{@{}l@{}}\textbf{Familiarity}\\\textbf{with LLMs}\end{tabular}} & Often use it                                       &      \multicolumn{1}{c}{2}     &     \multicolumn{1}{c}{1}    &      \multicolumn{1}{c}{2}  &   \multicolumn{1}{c}{1}  \\
& Occasionally use them                             &      \multicolumn{1}{c}{2}     &     \multicolumn{1}{c}{2}    &      \multicolumn{1}{c}{2}  &   \multicolumn{1}{c}{4}  \\
& Used them once or twice                             &      \multicolumn{1}{c}{2}     &     \multicolumn{1}{c}{4}    &      \multicolumn{1}{c}{2}  &   \multicolumn{1}{c}{2}  \\
& Never heard of/used them  &      \multicolumn{1}{c}{6}     &     \multicolumn{1}{c}{5}    &      \multicolumn{1}{c}{6}  &   \multicolumn{1}{c}{5}   \\ \arrayrulecolor{tablegrayline} \hline
\textbf{Participants} &                         Total      &    \multicolumn{1}{c}{12}                &  \multicolumn{1}{c}{12}             &  \multicolumn{1}{c}{12}                  &  \multicolumn{1}{c}{12} \\\arrayrulecolor{black}\hline
\end{tabular}%
}

\label{tab:demographics}
\end{table*}

\subsection{Data Analysis}
We collected rich dialogue data and valuable user subjective evaluation feedback. We performed both quantitative and qualitative analysis to examine chatbots' conversation styles, the slot filling performance, and participants' experiences with our chatbots. For each dialogue, we calculated commonly used descriptive metrics such as the number of turns and the average word counts per turn, which we report in \autoref{sec:results:descriptive}.

\ipstart{Slot Filling Performance} Our study aimed to investigate whether LLMs can drive chatbots to effectively ask defined questions and collect desired information specified in Table~\ref{tab:slots}. To calculate the amount of information that can be obtained by our chatbots, one researcher \textit{manually} inspected and determined whether each of the pre-defined information slots could be extracted from collected dialogues. More specifically, for \textit{sleep quality} and \textit{productivity rate}, which were specified as a scale of 1 to 10, we marked the slot as filled only if a numerical value (\eg, 9) rather than a vague phrase (\eg, good sleep) was given. For \textit{feelings after eating} in Food, we treated the slot to be filled if feelings regarding one or more meals were covered. We report the analysis of slot filling rate in \autoref{sec:results:slot}. Based on the binary coding, we calculated the \textbf{slot filling rate}: the ratio of the number of information slots extracted from the dialogue against the total number of slots. We use the slot filling rate to infer the data collection performance of chatbots.

\ipstart{Dialogue Acts and User/Chatbot Behaviors} To understand the conversational behaviors of the chatbots, we coded \textit{dialogue act} for each turn of conversations. Referring to some existing taxonomies of dialogue acts~\cite{shi2020effects, traum200020, welivita2020taxonomy}, three researchers independently coded one participant's dialogues (132 turns; 1.8\%) to identify emerging dialogue acts. Additionally, researchers labeled chatbot turns that did not fit in the conversation context or originated from the inherent artifacts of an LLM. We resolved discrepancies in coding and developed the first version of codebook with three dimensions of codes: (1)~essential acts and (2)~empathy \& engagement behaviors, and (3)~problematic chatbot turns. Then two researchers reiterated the independent coding of four other participants' dialogues (1 participant from each condition, 32 dialogues in total) with the codebook. The two researchers resolved discrepancies through multiple sessions of discussion until their inter-rater reliability (Cohen's Kappa) reached 0.96 for essential acts and 0.935 for empathy \& engagement behaviors. Compared to these dialogue acts, the occurrence of errors was sparse. Hence, the two researchers discussed the entire problematic turns coded by each other together and reached the full agreement. With the finalized codebooks~(See \autoref{tab:essential_components}, \ref{tab:empathy}, and \ref{tab:errors}), the first author coded the rest of the data.
As a result, each turn was classified as one of the essential acts---\textit{greeting}, \textit{task opening}, \textit{required question/answers (RQ/RA)}, \textit{secondary question/answers (SQ/SA)}, \textit{statement}, and \textit{closing}. We assigned the most prominent act to turns consisting of multiple sentences. Independent of essential acts, we  multi-coded each turn with the empathy \& engagement behaviors described in \autoref{tab:empathy}. For example, to a general compliment \symbolbot{}~``\textit{That's great}'' (\textbf{Statement}), we assigned only the \textbf{Appreciating} behavior, whereas we also treated \symbolbot{}~``\textit{That's great to hear that your legs are feeling stronger!}'' (\textbf{Statement}) to be both \textbf{Acknowledging} and \textbf{Appreciating} as the compliment directly addressed to the user input. We were interested in such acknowledging behaviors because \textit{specificity} was an important indicator of the capability of open-domain chatbots~\cite{adiwardana2020towards}.

\ipstart{Statistical Analysis} To understand the impact of the study factors, including prompt design, conversation topic, and the conversation path, to the chatbots' slot filling performance and conversational flows, we used \textit{mixed-effect models} because these models can handle unbalanced data repeatedly measured from the same participants~\cite{Pinheiro2000MixedEffects}. For each dialogue metric we want to assess, we fitted a mixed-effect model that predicts the metric, treating each dialogue as a data point. Starting from a full model containing participants as a random effect and the four main study factors--information format, personality modifier, topic, and path--and their interactions as fixed effects, we performed the step-wise backward elimination removing variables not significantly contributing the model, through Maximum-likelihood tests. For significant variables, we performed post-hoc pairwise comparisons of the least-squared means (LSM) of the metric using \texttt{emmeans}~\cite{emmeans} package in R.

\ipstart{Subjective Feedback}
To assess the difference among the experimental conditions, we conducted Kruskal-Wallis tests over the four rating questions. We also referenced the open-ended feedback from when interpreting the participants' reactions to specific phenomena of the conversations.
\section{Results}

In this section, we report the results of our study in six parts.
In \autoref{sec:results:descriptive}, we provide an overview of the dialogue dataset we collected. 
In \autoref{sec:results:slot}, we report the data collection performance of our chatbots and factors that impact the performance. 
In \autoref{sec:results:essential}, we report the types of the essential dialogue acts and assess how the prompt design and other factors impact the dialogue acts and, in turn, the data collection performance. 
In \autoref{sec:results:empathy}, we report the types of the empathetic and engaging behaviors of chatbots and assess how the prompt design and other factors impact such behaviors of the chatbots.
In \autoref{sec:results:errors}, we explore the problematic chatbot utterances mainly caused by the erroneous behaviors of a large language model. Lastly, in \autoref{sec:results:subjective}, we report on participants' subjective evaluation from the exit surveys.

\subsection{Descriptive Statistics}\label{sec:results:descriptive}

\begin{table*}[b]    
    \small\sffamily
			\def\arraystretch{1.3}
		    \setlength{\tabcolsep}{0.5em}
		    \centering
\caption{Descriptive statistics of our dialogue dataset aggregated by four prompt designs.}
\label{tab:descriptive_stats}
\begin{tabular}{|l!{\color{lightgray}\vrule}cccc|}
\hline
\rowcolor{tableheader}
                                   & \textbf{SP} & \textbf{SN}     & \textbf{DP} & \textbf{DN}     \\ \hline
Total number of dialogues (turns)             & 91 (1,638)    & 96 (1,889)    & 95 (1,941)    & 92 (1,975)    \\ \arrayrulecolor{tablegrayline}\hline
Average no. of turns per dialogue (range)  & 18.0 (7--45)  & 19.7 (3--57)  & 20.4 (7--75)  & 21.47 (3--53) \\ \hline
Average no. of words per dialogue  & 212.3 & 240.8 & 321.7 & 277.1 \\ \hline
Average no. of chatbot/user words per turn  & 17.4 / 4.9 & 17.8 / 4.8  & 23.4 / 7.5  & 19.2 / 5.5  \\ \hline
Percentage of organically ended conversations & 71.4\%   & 76.0\%   & 77.9\%  & 78.2\%  \\ \hline
Percentage of erroneous turns & 3.1\%    & 4.3\%   & 3.0\%  & 3.6\%  \\ 
\arrayrulecolor{black}\hline
\end{tabular}
\end{table*}

From 48 participants, we collected 374 dialogues (7,442 turns in total): 91 from SP; 96 from SN; 95 from DP, and 91 from DN. Regarding the conversation topic, 94, 91, 94, and 95 dialogues were from Sleep, Work, Exercise, and Food Intake, respectively. Eight participants missed one dialogue per each and one missed two, mainly due to temporary server issues or accidental skips.

Prompt designs impacted the word lengths and the number of turns of chatbots. \autoref{tab:descriptive_stats} summarizes the number of turns and word counts by prompt design. The average number of turns per dialogue is around 20 with more average turns under the two descriptive conditions (DP and DN). The maximum number of dialogue turns is 75 under the DP condition (only 1 dialogue). In terms of word counts, dialogues under the descriptive conditions (DP, DN) had more words than those under the structured conditions (SP, SN): both chatbots and participants uttered more words under the descriptive conditions. The DP condition, in particular, leads to the most number of words of dialogues.

\subsection{Slot Filling Rate}\label{sec:results:slot}

\begin{table*}[b]
    \small\sffamily
	\def\arraystretch{1.1}\setlength{\tabcolsep}{1em}
		    \centering
\caption{The slot filling rate (and $SD$) by topic and condition.}
{%
\begin{tabular}{|l|llll!{\color{lightgray}\vrule}l|}
\hline
\rowcolor{tableheader}
            & \textbf{Sleep} & \textbf{Work} & \textbf{Food Intake} & \textbf{Exercise} & \textbf{Total}\\ \hline
\textbf{SP} &   0.83 (0.29)    &  0.71 (0.25)    &  0.85 (0.18)    &   0.93 (0.14)  &  0.83 (0.23)   \\ \arrayrulecolor{tablegrayline}\hline
\textbf{SN}  &   0.75 (0.30)    &  0.64 (0.32)    &  0.80 (0.33)    &   0.88 (0.20)  &  0.77 (0.30)     \\ \hline
\textbf{DP} &   0.67 (0.24)    &  0.67 (0.28)    &  0.72 (0.32)    &   0.82 (0.20)  &  0.72 (0.27)     \\ \hline
\textbf{DN} &   0.85 (0.16)    &  0.83 (0.24)    &  0.75 (0.24)    &   0.91 (0.16)  & 0.83 (0.21)\\ \hline
\textbf{Total} & \textbf{0.77 (0.26)} & \textbf{0.71 (0.28)}  & \textbf{0.78 (0.28)} &  \textbf{0.88 (0.18)} & \textbf{0.79 (0.26)}\\ \arrayrulecolor{black}\hline
\end{tabular}%
}
\label{tab:completion_rate}
\end{table*}

\begin{figure*}[b]
    \begin{flushleft}
    \siglegend
    \end{flushleft}
    \centering
    \includegraphics[width=0.95\textwidth]{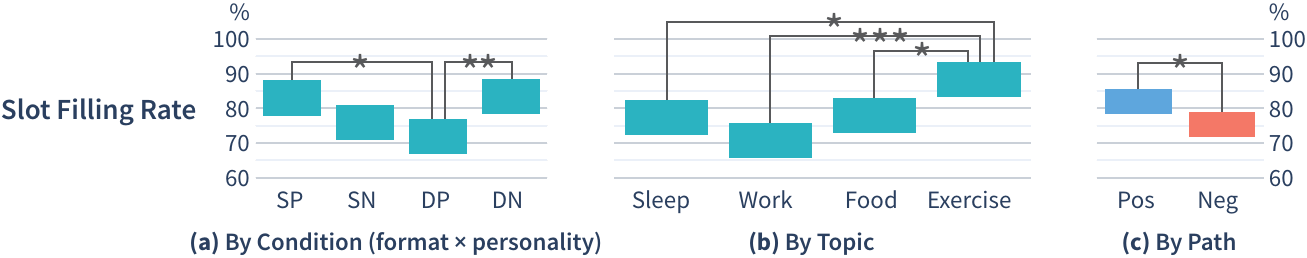}
    
    \labelphantom{fig:slot_emmeans:condition}
    \labelphantom{fig:slot_emmeans:topic}
    \labelphantom{fig:slot_emmeans:path}
    
    \caption{95\% confidence intervals of slot filling rate by variables with a significant effect: (a) the combination of the information format and personality modifier represented as study condition; (b) topic; and (c) the conversation path. The asterisks with arms indicate significance between the connected categories. (Refer to Appendix~\ref{appendix:stats:slot} for model details and statistics.)}
    \label{fig:slot_emmeans}
\end{figure*}

Prompt designs significantly impacted the slot filling performance of chatbots. \autoref{tab:completion_rate} summarizes the average slot filling rates of chatbots by conditions and topics. On average, all chatbots have reached over 70\% slot filling rates. The dialogues in the SP-Exercise condition had the highest rate (93\%) and those in the SN-Work condition had the lowest rate (64\%).
The maximum-likelihood test revealed that there was no significant random effect of participants, indicating that participants have little impact on chatbots' data collection performance. On the other hand, there were significant random effects of the topics($p<.0001$), the conversation paths ($p=.01$), and the interaction between the information formats and personality modifiers ($p<.001$).
\autoref{fig:slot_emmeans} shows the significance over the 95\% confidence intervals of the slot filling rate in each category of the significant variables. The dialogues in DP condition had significantly lower rates than those in SP ($p=.01$) and DN ($p=.008$). This suggests that the personality modifier impacted chatbots differently: with the modifier, chatbots with the structured prompt yield higher rates whereas chatbots with the descriptive format yield higher rates without the modifier~(See \autoref{fig:slot_emmeans:condition}). In terms of topic, Exercise dialogues had the highest rate of 88.4\%, which was significantly higher than those in the other three topics: Sleep ($p=.01$), Work ($p<.0001$), and Food ($p=.02$) (See \autoref{fig:slot_emmeans:topic}). Lastly, dialogues in the Positive path had significantly higher rates than those in the Negative path ($p=.01$) (See \autoref{fig:slot_emmeans:path}).

\begin{figure*}[h]
\includegraphics[width=\textwidth]{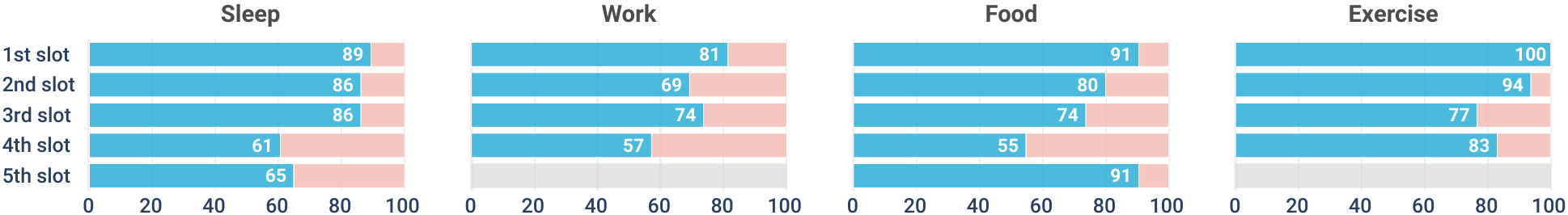}
\caption{Breakdowns of the percentage of filled slots by the order of questions for each topic. Work and Exercise consist of four slots.}
\label{fig:slot_by_question}
\end{figure*}

As seen in \autoref{fig:slot_by_question}, there is a general trend that slots specified earlier in prompts were more likely to be covered by chatbots. For example, the first slots in all topics were covered in 90.3\% of the dialogues, but the last specified slots in Sleep (sleep quality) and Work (what went well) were omitted around 40\% of the dialogues. Interestingly, the last specified slot of Food (feelings after eating) was diligently covered: chatbots often asked how participants felt after talking about each meal rather than asking their feelings once towards the end.

\subsection{Essential Dialogue Acts}\label{sec:results:essential}

%

\begin{table*}[b]
    \footnotesize\sffamily
	\def\arraystretch{1.2}\setlength{\tabcolsep}{0.25em}
		    \centering
\caption{Summary of essential dialogue acts with the ratio of the occurring turns per dialog (turn ratio) by condition, brief description, and exemplar turns (\symboluser{}: user turns, \symbolbot{}: chatbot turns).}
\label{tab:essential_components}
\begin{tabular}{|p{0.12\textwidth}|cccc|p{0.25\textwidth}!{\color{lightgray}\vrule}p{0.38\textwidth}|}
\hline
\rowcolor{tableheader}
 \textbf{Dialogue Act} & \multicolumn{4}{c|}{\textbf{Turn Ratio (\%)}} & \textbf{Description} & \textbf{Examples} \\
  \rowcolor{tableheader}  & \textbf{SP} & \textbf{SN} & \textbf{DP} & \textbf{DN} & & \\\hline 
  
\textbf{Greeting} & 12.40 & 12.94 & 10.67 & 10.50 & Initiation of a conversation. &
  \parbox[t]{0.45\textwidth}{\symbolbot{} \textit{How's your day so far?}\\ \symboluser{} \textit{I feel refreshed and recharged.}\vspace{0.5mm}} \\
  \arrayrulecolor{tablegrayline}\hline
  
  \textbf{Task opening} & 1.92 & 1.56 & 0.54 & 0.04 &  \parbox[t]{0.30\textwidth}{General questions that bring \\up the conversation topic.} &
  \symbolbot{} \parbox[t]{0.32\textwidth}{\textit{How was your work and productivity\\\hspace{5mm}yesterday?}} \\\hline
  
  \textbf{Required question} & 24.07 & 21.36 & 17.57 & 20.97 & \multirow{2}{*}{\parbox[t]{0.30\textwidth}{Questions and answers that\\are directly related to the\\specified information slots.}}  &
  \multirow{2}{*}{\parbox[t]{0.45\textwidth}{\symbolbot{} \textit{What was your lunch yesterday?} (RQ)\\ \symboluser{} \textit{I had pork barbeque.} (RA)\vspace{0.5mm}}} \\ \cline{1-5}
  \parbox[t]{0.26\textwidth}{\textbf{Required\\answer}\vspace{1mm}} & 23.74 & 19.91 & 17.35 & 20.75 & & \\ \hline
  
 \textbf{Secondary question} & 10.28 & 10.79 & 13.87 & 13.91 & \multirow{2}{*}{\parbox[t]{0.26\textwidth}{Questions and answers that\\are usually follow-ups and\\not specific to the slots.}} &
  \multirow{2}{*}{\parbox[t]{0.45\textwidth}{\symbolbot{} \textit{What did you put on your toast?} (SQ)\\ \symboluser{} \textit{I put strawberry jam on it.} (SA)\vspace{0.5mm}}} \\ \cline{1-5}
    \textbf{Secondary answer} & 9.12 & 9.64 & 13.08 & 12.71 & & \\ \hline
  
  \textbf{Statement} & 14.22 & 19.54 & 22.85 & 16.94 & \parbox[t]{0.30\textwidth}{Non-Q\&A messages such as\\commenting or summarizing.} &
  \parbox[t]{0.37\textwidth}{\symbolbot{} \textit{It sounds like you had a great night's sleep!}\\ \symboluser{} \textit{I did! Not sure why though.}\vspace{0.5mm}} \\\hline
  
  \textbf{Closing} & 3.88 & 4.31 & 3.96 & 4.14 & Farewell or ending messages. &
  \symbolbot{} You're welcome. Have a great day! \\ 
\arrayrulecolor{black}\hline
\end{tabular}
\end{table*}

To further understand how chatbots powered by different prompt designs talk, we categorized conversation turns into dialogue acts. We provide the summary of essential dialogue acts and their distributions in \autoref{tab:essential_components}
Here, we report chatbots' essential acts regarding question/answering and non-question statements.

\subsubsection{Required and Secondary Questions} 
We identified two types of questions that the chatbots asked: required questions (RQ) and secondary questions (SQ). The RQs were directly related to the specified information slots, whereas SQs were not directly related to the information slots but rather follow-up details or elaboration. Despite being relevant to the conversation topic, SQs sometimes caused the conversation to digress. Although not very common (95 out of 1,029 SQ turns in total; 9.3\%), participants also asked questions to the chatbot, which were all categorized as SQ/SA. The majority of collected dialogues consisted of question/answering: Overall, 4,879 out of 7,442 turns ($avg.$ 64.72\% of turns per dialogue; $min=10.67\%$, $max=93.54\%$) were classified as RQ, RA, SQ, or SA (See \autoref{tab:essential_components}).

\begin{figure*}[b]
    \begin{flushleft}
    \siglegend
    \end{flushleft}
    \centering
    \includegraphics[width=\textwidth]{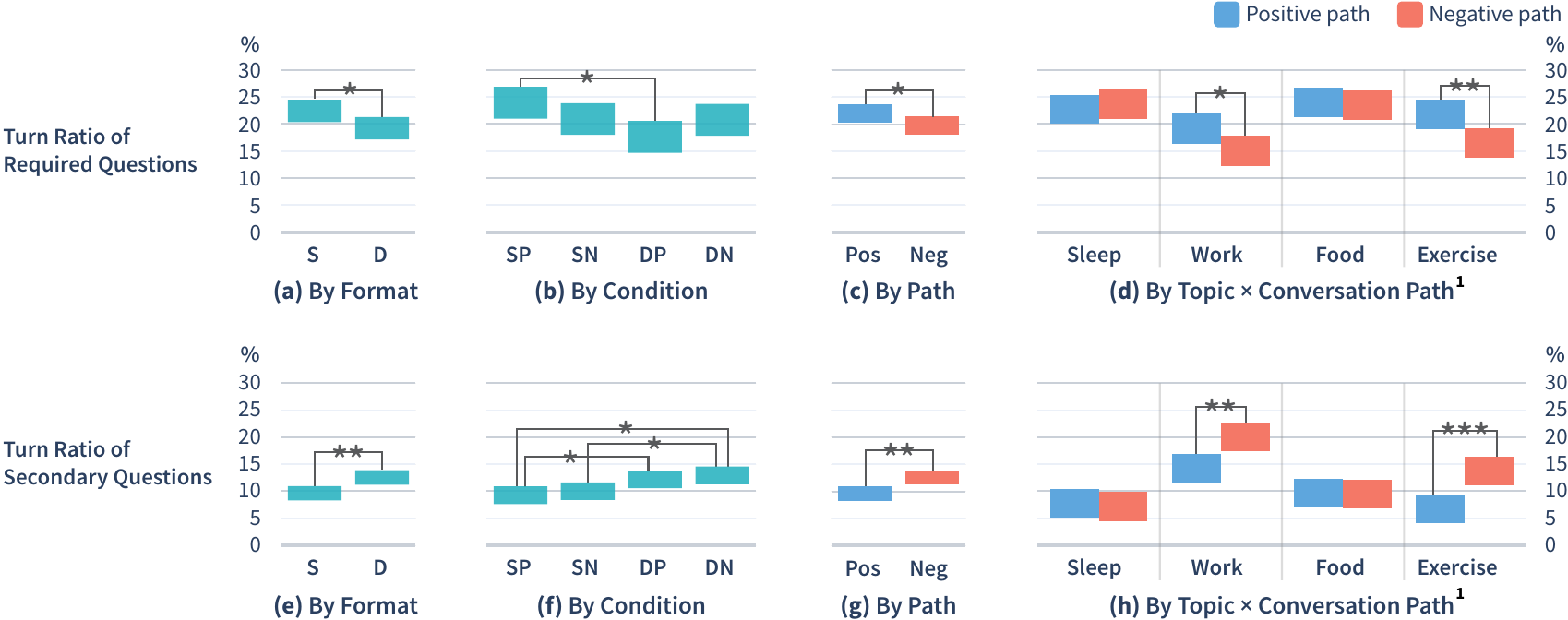}
    \labelphantom{fig:questions_emmeans:rq_format}
    \labelphantom{fig:questions_emmeans:rq_condition}
    \labelphantom{fig:questions_emmeans:rq_path}
    \labelphantom{fig:questions_emmeans:rq_topicpath}
    \labelphantom{fig:questions_emmeans:sq_format}
    \labelphantom{fig:questions_emmeans:sq_condition}
    \labelphantom{fig:questions_emmeans:sq_path}
    \labelphantom{fig:questions_emmeans:sq_topicpath}
    
    \begin{flushright}
    \footnotesize{$^1$Excluded pairs across different topics from the pairwise comparison.}
\end{flushright}
    
    \caption{95\% confidence intervals of the turn ratios of RQ (top; a--d) and SQ (bottom; e--h) by variables with a significant effect: The asterisks with arms indicate significance between the connected categories. Note that for (d) and (h) we did not display the significance across topics. (Refer to Appendices \ref{appendix:stats:rq} and \ref{appendix:stats:sq} for model details and statistics.)}
    \label{fig:questions_emmeans}
\end{figure*}

We first investigated the impact of prompt designs on the chatbot-spoken RQ and SQ turn ratios using two mixed-effect models with each turn ratio as a dependent variable, respectively. 
\autoref{fig:questions_emmeans} shows the 95\% confidence intervals of the chatbot-spoken RQ and SQ turn ratios by four study factors (information format, personality modifier, topic, and path). The structured format significantly increased the RQ turn ratio ($p=.03$; see \autoref{fig:questions_emmeans:rq_format}) but decreased the SQ turn ratio ($p=.002$; see \autoref{fig:questions_emmeans:sq_format}). On the other hand, the personality modifier did not impact either RQ ($p=.94$) nor SQ ($p=.43$) turn ratios and made no difference within the same information format (See \autoref{fig:questions_emmeans:rq_condition} and \ref{fig:questions_emmeans:sq_condition}). 

In terms of the conversation path, we find that, overall, the Positive path led to a higher RQ turn ratio ($p=.01$; see \autoref{fig:questions_emmeans:rq_path}) and a lower SQ turn ratio ($p=.002$; see \autoref{fig:questions_emmeans:sq_path}). However, under different topics, the conversation path had different impacts on RQ and SQ ratios. As seen in \autoref{fig:questions_emmeans:rq_topicpath}, the Positive path increased the RQ turn ratio only in the Work ($p=.02$) and Exercise ($p=.003$) dialogues and also decreased the SQ turn ratio in the same topics (Work: $p=.003$ and Exercise: $p<.001$).

As discussed above, prompt designs, topics and conversation paths have significant impacts on chatbots' question-asking behaviors. The RQ and SQ ratios further impacted the slot filling rate. We ran the maximum-likelihood tests with two mixed-effect models fitting the slot filling rate, one with the chatbot-spoken RQ turn ratio (\ie, the ratio of the turns classified as RQ in a dialogue) as a fixed effect and the other with the SQ turn ratio, both with participants as a random effect. We found that the RQ turn ratio was positively correlated with the slot filling rate, whereas the SQ turn ratio was negatively correlated with it: $\beta=1.08$, $SE=0.12$, $t(347.25)=8.80$, $p < .0001$ for RQ and $\beta=-0.77$, $SE=0.13$, $t(370.06)=-6.12$, $p < .0001$ for SQ.

In summary, our results suggest that chatbots with a Descriptive information format tend to ask more secondary questions, and negative answers of participants also naturally elicit more secondary questions (\eg, \symbolbot{}~``\textit{I'm sorry to hear that you didn't workout yesterday. May I ask why?}'').

\subsubsection{Non-question Statements}
Statements from chatbots usually consist of chitchats, comments and suggestions that do not fall within the question/answering (see \autoref{tab:essential_components} for references) and participants also tend to respond to statements with other statements. Usually, the participants' statements did not contain direct clues about the desired information slots. Overall, 1,482 out of 7,442 turns ($avg.$ 18.45\% per dialogue) were classified as Statement (See \autoref{tab:essential_components}). Among the four conditions, dialogues of DP have the highest ratio of statements (24.99\%) and SP had the lowest ratio (15.32\%). 

\subsection{Empathy \& Engagement Behaviors}\label{sec:results:empathy}

%

\begin{table*}[b]
    \footnotesize\sffamily
	\def\arraystretch{1.1}\setlength{\tabcolsep}{0.15em}
		    \centering
\caption{Summary of empathy \& engagement behaviors with turn ratio by condition, brief description, and exemplar turns. Note that the behaviors are multi-coded.}
\label{tab:empathy}
\begin{tabular}{|p{0.15\textwidth}|cccc|p{0.24\textwidth}!{\color{lightgray}\vrule}p{0.40\textwidth}|}
\hline
\rowcolor{tableheader}
 \textbf{Behavior} & \multicolumn{4}{c|}{\textbf{Turn Ratio (\%)}} & \textbf{Description} & \textbf{Examples} \\
  \rowcolor{tableheader} \textbf{Category}  & \textbf{SP} & \textbf{SN} & \textbf{DP} & \textbf{DN} & & \\\hline 
  
\textbf{Acknowledging} & 18.22 & 18.91 & 26.24 & 22.04 & \parbox[t]{0.22\textwidth}{Directly referring to what \\the other said.} &
  \symbolbot{} \parbox[t]{0.36\textwidth}{ \textit{That's great to hear that your legs are\\feeling stronger!}\vspace{0.5mm}} \\\arrayrulecolor{tablegrayline}\hline
  
\textbf{Appreciating} & 9.04 & 7.36 & 11.18 & 9.86 & Complimenting the other. &
  \symbolbot{} \textit{That's terrific!} \\ \hline
  
\textbf{Sympathizing} & 3.66 & 2.79 & 4.33 & 1.74 & Sympathizing with the other. &
  \symbolbot{} \textit{I'm sorry to hear that. What's been going on?} \\ \hline

\textbf{Thanking} & 5.16 & 6.74 & 7.05 & 7.26 & Being grateful to the other. &
  \symboluser{} \textit{I feel nice. Thanks for asking.} \\\hline
  
\parbox[t]{0.15\textwidth}{\textbf{Advice/\\suggesting}} & 2.54 & 3.07 & 4.81 & 2.55 & Giving advice or suggestion. &
  \symbolbot{} \parbox[t]{0.36\textwidth}{ \textit{I can give you some recommendations on\\exercises that will help you grow your\\adductor muscles.}\vspace{1mm}} \\\hline
  
\parbox[t]{0.15\textwidth}{\textbf{Rejecting/\\disagreeing}} & 0.57 & 0.66 & 0.40 & 0.66 & Rejecting or disagreeing with the other. &
  \symboluser{} \textit{But the weather needs to be good for walking.} \\
  \arrayrulecolor{black}\hline
\end{tabular}
\end{table*}

\begin{figure*}[t]
    \begin{flushleft}
    \siglegend
    \end{flushleft}
    \centering
    \includegraphics[width=\textwidth]{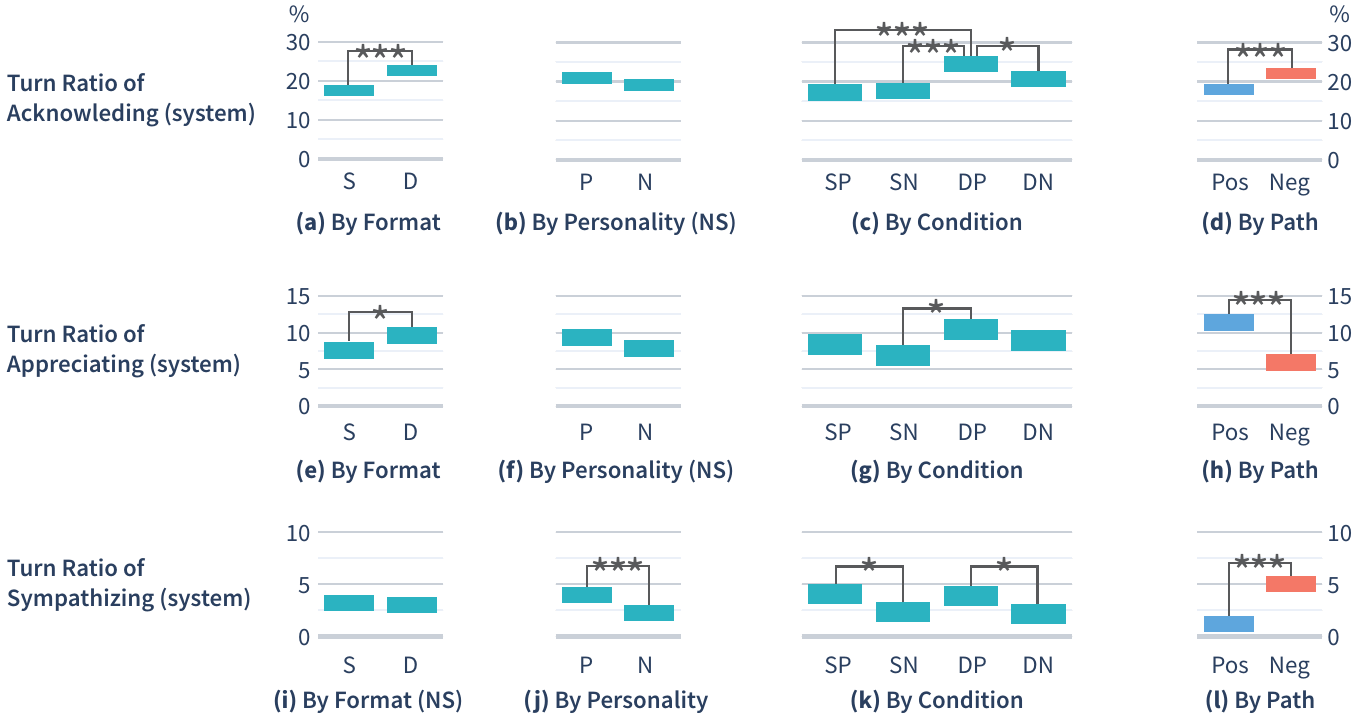}
    
    \labelphantom{fig:empathy_emmeans:ack:format}
    \labelphantom{fig:empathy_emmeans:ack:personality}
    \labelphantom{fig:empathy_emmeans:ack:condition}
    \labelphantom{fig:empathy_emmeans:ack:path}
    \labelphantom{fig:empathy_emmeans:appr:format}
    \labelphantom{fig:empathy_emmeans:appr:personality}
    \labelphantom{fig:empathy_emmeans:appr:condition}
    \labelphantom{fig:empathy_emmeans:appr:path}
    \labelphantom{fig:empathy_emmeans:symp:format}
    \labelphantom{fig:empathy_emmeans:symp:personality}
    \labelphantom{fig:empathy_emmeans:symp:condition}
    \labelphantom{fig:empathy_emmeans:symp:path}
    
    \caption{95\% confidence intervals of the chatbot turn ratios for Acknowledging (a--d), Appreciating (e--h), and Sympathizing (i--l) behaviors by information format, personality modifier, study condition (combinations of format and personality modifier), and conversation path. Variables that are not significant are marked as `NS.' (Refer to Appendix \ref{appendix:stats:empathy} for model details and statistics.)}
    \label{fig:empathy_emmeans}
\end{figure*}

\autoref{tab:empathy} summarizes the empathy \& engagement categories and their turn ratios by experimental condition. The majority of these behavior categories were coded to the chatbot turns---1,992 chatbot turns and 403 user turns were coded with one or more behavior categories---partly because participants uttered less words than chatbots (see \autoref{tab:descriptive_stats}) and chatbots usually led the conversation while participants simply responded. \textbf{Acknowledging} was the most common empathy behavior (see \autoref{tab:empathy}) as chatbots often referred to what participants previously said in generated messages. Also, we found that our chatbots often \textbf{appreciated} participants' accomplishment (\eg, taking good sleep, managed to exercise) or \textbf{sympathized} participants when they reported negative outcomes (\eg, poor sleep quality, failed at work).

To investigate how the four study factors impact the empathetic behaviors of chatbots, we analyzed three mixed-effect models with the chatbot turn ratios of Acknowledging, Appreciating, and Sympathizing behaviors as a dependent variable, respectively.
\autoref{fig:empathy_emmeans} shows the 95\% confidence intervals of turn ratios of the three behavior categories estimated against the study factors. The information format significantly influenced the Acknowledging and Appreciating turn ratios: Dialogues in the Descriptive format had higher ratios of the Acknowledging ($p<.0001$; see \autoref{fig:empathy_emmeans:ack:format}) and Appreciating ($p=.01$;  see \autoref{fig:empathy_emmeans:appr:format}) turns. Personality modifier did not solely impact these two behaviors but it influenced in conjunction with the information format (See \autoref{fig:empathy_emmeans:ack:condition} and \ref{fig:empathy_emmeans:appr:condition}). However, the personality modifier in the prompt led chatbots to produce significantly more Sympathizing turns ($p=.002$; see \autoref{fig:empathy_emmeans:symp:personality}). Besides the prompt design, the conversation path strongly influenced all three empathetic behaviors: The Positive path led to higher turn ratio of Appreciating ($p<.0001$; see \autoref{fig:empathy_emmeans:ack:path}) whereas The Negative path led to higher Acknowledging ($p<.001$; see \autoref{fig:empathy_emmeans:appr:path}) and Sympathizing ($p<.0001$; see \autoref{fig:empathy_emmeans:symp:path}) turn ratios. 

\subsection{Problematic Chatbot Turns and User Responses}
\label{sec:results:errors}
\begin{table*}[b]
    \footnotesize\sffamily
	\def\arraystretch{1.2}\setlength{\tabcolsep}{0.15em}
		    \centering
\caption{Categories of the chatbots' erroneous turns with turn ratio, brief descriptions, and examples.}
\label{tab:errors}
\begin{tabular}{|p{0.14\textwidth}|p{0.10\textwidth}!{\color{lightgray}\vrule}p{0.24\textwidth}!{\color{lightgray}\vrule}p{0.49\textwidth}|}
\hline
\rowcolor{tableheader}
\textbf{Category} & \textbf{Turn Ratio} & \textbf{Description} & \textbf{Example} \\ \hline
\textbf{Incorrect phrases} & 0.79\% & \parbox[t]{0.24\textwidth}{Messages with symbols or \\extra predictions.\vspace{0.5mm}} & \begin{tabular}{@{}l@{}l@{}}\symbolbot{}\textit{Workout duration -\textgreater{} 45 minutes} (1) \\ \symbolbot{} \textit{It took me a while to fall asleep.} (2) \vspace{-2.5mm} \end{tabular}  \\ \arrayrulecolor{tablegrayline}\hline

 \hline

\textbf{Self-talk}       & 1.1\% & \parbox[t]{0.24\textwidth}{First-person narratives or \\commentary messages.\vspace{0.5mm}} & \begin{tabular}{@{}l@{}l@{}}\symbolbot{} & \parbox[t]{0.38\textwidth}{\textit{The customer's fitness concern is that they are feeling very tired after their workout.}\vspace{-2.5mm}} \end{tabular} \\ \hline

\textbf{Repetition}      & 3.8\% & Repeating the same or similar utterances multiple times. & Refer 
 to~\autoref{tab:repetition_eg}. \\ \hline
\textbf{Miscellaneous}           & 0.9\% & \parbox[t]{0.24\textwidth}{Other miscellaneous errors.\\ \vspace{0.5mm}}      & \begin{tabular}{@{}l@{}l@{}}\symbolbot{}\textit{    (no response)} (1) \\ \symbolbot{} \textit{Other than work, what were your concerns at work?} (2) \vspace{-1.5mm} \end{tabular} \\ 
\arrayrulecolor{black}\hline
\end{tabular}
\end{table*}
In total, 6.7\% of the chatbot turns (257 out of 3,916) were tagged erroneous and the four categories of erroneous turns are summarized in \autoref{tab:errors}. These erroneous turns sometimes led to the unorganic termination of the conversation (\ie, participants ended the conversation before or without natural Closing messages). In the following, we cover these error types in detail.

\ipstart{Incorrect phrases} In structured prompts (SP, SN), we used a symbol ``\texttt{->},'' a commonly-used delimiter for key-value pairs in LLMs, to specify the information slots. This caused GPT-3 to expose such an information structure to the output as an artifact in 16 turns of messages, all of which were generated by the structured format. In the example from~\autoref{tab:errors}, GPT-3 even incorrectly ``predicted'' the slots (45 minutes, example (1) in~\autoref{tab:errors}) altogether with symbols. Other times, GPT-3 also erroneously predicted the answers for participants (example (2)~\autoref{tab:errors}). Particularly, we identified 6 instances where GPT-3 predicted the user response and appended an extra \symboluser{}turn to the generated chatbot turns (3 turns in descriptive groups and 3 turns in structured groups). \majorrev{While the exposure of the symbol was caused by the structured prompt design, other incorrect phrases were mostly caused by the random nature of LLMs~\cite{brown2020language}.}

\ipstart{Self-talk} GPT-3 sometimes generated turns in a first-person narrative or not directed to participants, which looked quite similar to the ``self-talk'' of humans. In \autoref{tab:self_error}, for example, Turn 03 is obviously not directed to participants. In our dataset, less than 1.1\% of chatbot turns were self-talk (SP: 8, SN: 27, DP: 1, DN: 9). Participants who encountered these ``self-talk'' commented that these messages were ``awkward'' (P42), ``strange'' (P24), and ``confusing''(P23). However, we found that participants always attempted to continue the conversation by following the self-talk and tried to resolve the errors (See Turn 04 in~\autoref{tab:self_error}). \majorrev{We used the first-person narrative in prompts as it appeared to make generated chatbots talk more like humans; however, the first-person narrative might negatively lead to the ``self-talk'' in conversations.}

\begin{fragment}
    \def\arraystretch{1.1}\setlength{\tabcolsep}{0.35em}
    \small\sffamily\centering
    \caption{Dialogue snippets of self-talk from P43.}
    \label{tab:self_error}
    \begin{tabularx}{\columnwidth}{lll}
    \hline
    01 &
      \symbolbot{} & \textit{Do you have any fitness concerns that you would like to address?} \\
      \arrayrulecolor{tablegrayline}\hline
    02 &
      \symboluser{} &
      \textit{nope not at the moment.} \\\hline
    03 &
      \symbolbot{} &
      \begin{tabular}[c]{@{}l@{}}\textcolor{magenta}{\textit{\textbf{Based on the customer's responses, it seems that they did not have a very active day}}}\\\textcolor{magenta}{\textbf{\textit{yesterday. I would encourage them to try and get some form of physical activity in}}} \\ \textcolor{magenta}{\textit{\textbf{today, even if it is just a short walk.}}}\end{tabular} \\\hline
    04 &
      \symboluser{} &
      \textit{yes I should probably try it tomorrow.}
      \\\arrayrulecolor{black}\hline
    \end{tabularx}

\end{fragment}

\ipstart{Repetition} 
We found that GPT-3 was susceptible to generate repetitive messages, either identically or linguistically repeating the previous chatbot turns. In total, 147 turns (3.8\% of chatbot turns) were labeled to be repetitive (SP: 11, SN: 39, DP: 48, DN: 50). Identically repetitive messages occurred to 23 participants (SP: 3, SN:5, DP: 6, DN: 9) in 31 dialogues (SP: 3, SN: 9, DP: 8, DN:10). Among the four topics, work (14 dialogues) tended to have more identically repetitive messages and exercise (2 dialogues) tended have fewer identically repetitive messages. However, these messages usually served as SQ/SA in conversations, hence they rarely influenced slot filling and data collection. Linguistically repetitive messages usually share similar wording or phrase structures. \autoref{tab:repetition_eg} presents an example dialogue snippet. At Turn 01, 03, 05, and 07, the chatbot always started with a similar phrase (\ie, \symbolbot{}``\textit{That's great ...}'') to compliment the participant then asked the participant a question started with ``\textit{can you}''. These linguistically repetitive messages were semantically correct and \textit{58.3\%} took place in the Acknowledging turns where the chatbot rephrased what the participant said progressed the conversation organically. However, linguistically phrased messages looked too similar, and they negatively impacted the user experience. A few participants suggested that the chatbots seemed to actually understand their responses, yet were using ``a sentence template'' (P29) to respond in a ``predefined ways'' (P13). On the other hand, repetition could also lead to ``dead loops'' of conversations. \majorrev{Repetitions were mostly caused by principles of LLMs. As the conversation history was used as the new prompt, GPT-3 tended to generate patternized texts, which led to further repetitions.}

\begin{fragment}
    \def\arraystretch{1.0}\setlength{\tabcolsep}{0.3em}
    \small\sffamily\centering
\caption{Dialogue snippet of linguistically repetitive utterances from P37.}
    \begin{tabularx}{\columnwidth}{lll}
    \hline
    01 & \symbolbot{} & \textit{That's a specific and achievable goal. Can you tell me more about what libraries you need to install?} \\\arrayrulecolor{tablegrayline}\hline
    02 & \symboluser{} & \textit{I don't know yet which libraries I need, but probably Pytorch, ...}                                    \\\hline
    03 & \symbolbot{} & \textit{\textcolor{magenta}{\textbf{That's a great goal. Can you}} research what libraries you need and then install them?}\\\hline
    04 & \symboluser{} & \textit{The libraries I need is provided in a file.}                                                           \\\hline
    05 & \symbolbot{} & \textit{\textcolor{magenta}{\textbf{That's great.}} So you have a list of the libraries you need. \textcolor{magenta}{\textbf{Can you install them?}}}                     \\\hline
    06 & \symboluser{} & \textit{I need to run a script that download and install these libraries.}                                     \\\hline
    07 & \symbolbot{} & \textit{\textcolor{magenta}{\textbf{That's a great goal. Can you}} walk me through the script?} \\\arrayrulecolor{black}\hline                                            
    \end{tabularx}
    \label{tab:repetition_eg}
\end{fragment}

\ipstart{Miscellaneous} There were 37 problematic chatbot turns that did not fall into the above categories. Among these turns, 19 turns were tagged to have context errors (\ie, chatbots did not grasp the context at all), 10 had semantic errors (\ie, messages that are not of human common sense), and one had both context and semantic errors. In \autoref{tab:context_error}, for example, the chatbot entirely missed that the participant said \symboluser{}``\textit{\textbf{1 hour} of cardio}.'' But the chatbot also attempted to resolve the contextual misunderstanding by apologizing after the participant corrected it. While chatbots in our study appeared to understand people pretty well in most cases, their responses with semantic errors could be quite wrong and amusing. For instance, when being asked about workout yesterday, one participant told the chatbot that they skipped. Instead of considering the participant did not workout, the chatbot considered skipping as a jumping workout and responded \symbolbot{}``\textit{Skipping is a type of cardiovascular exercise that can help to improve your heart health and endurance.}'' Besides, there were two instances when the chatbot failed to detect the ending of the conversation and restarted with the first slot question again, which, of course, led participants to abandon the conversation. \majorrev{These errors were rare but could also be attributed to the randomness nature of LLMs.} Lastly, our system went offline 5 times and caused chatbots to output empty messages, which was caused by the over-frequent API calls to OpenAI.

\ipstart{Terminating Conversations}
With the Closing turns, we found that 75.7\% of the conversations were organically ended. Among the four conditions, SP had the lowest percentage of naturally ended conversations (71.4\%) than other three conditions. For 91 non-organically ended conversations, participants abandoned 48.3\% of those conversations without encountering any obvious problematic errors. 19.8\% of prematurely ended conversations were caused by identically repetitive and another 19.8\% of were caused by linguistically repetitive messages. Lastly, the self-talk led to 3 conversations to end early, and context errors and system-offline caused the rest incomplete conversations.

\begin{fragment}
    \small\sffamily\centering
    \caption{Dialogue snippet of the context error from P6.}
    \label{tab:context_error}
    \begin{tabularx}{\columnwidth}{lll}
    \hline
    01 & \symbolbot{} & \textit{What workout did you do yesterday?}                                                                 \\\arrayrulecolor{tablegrayline}\hline
    02 & \symboluser{} & \textit{I did \textbf{1 hour} of cardio yesterday.}                                                                                                                                                            \\\hline
    03 & \symbolbot{} & \textit{\textcolor{magenta}{\textbf{How long did your workout last?}}}                                                                           \\\hline
    04 & \symboluser{} & \textit{Come on dude. I just told you I did 1 hour of cardio. How long is 1 hour?}
    \\\hline
    05 & \symbolbot{} & \begin{tabular}[c]{@{}l@{}}\textit{I'm sorry if I came across as rude. I was just trying to get some information from you about}\\\textit{your workout yesterday.}\end{tabular} \\\arrayrulecolor{black}\hline
    \end{tabularx}    
\end{fragment}

\begin{figure*}[t]
    \centering
    \includegraphics[width = \textwidth]{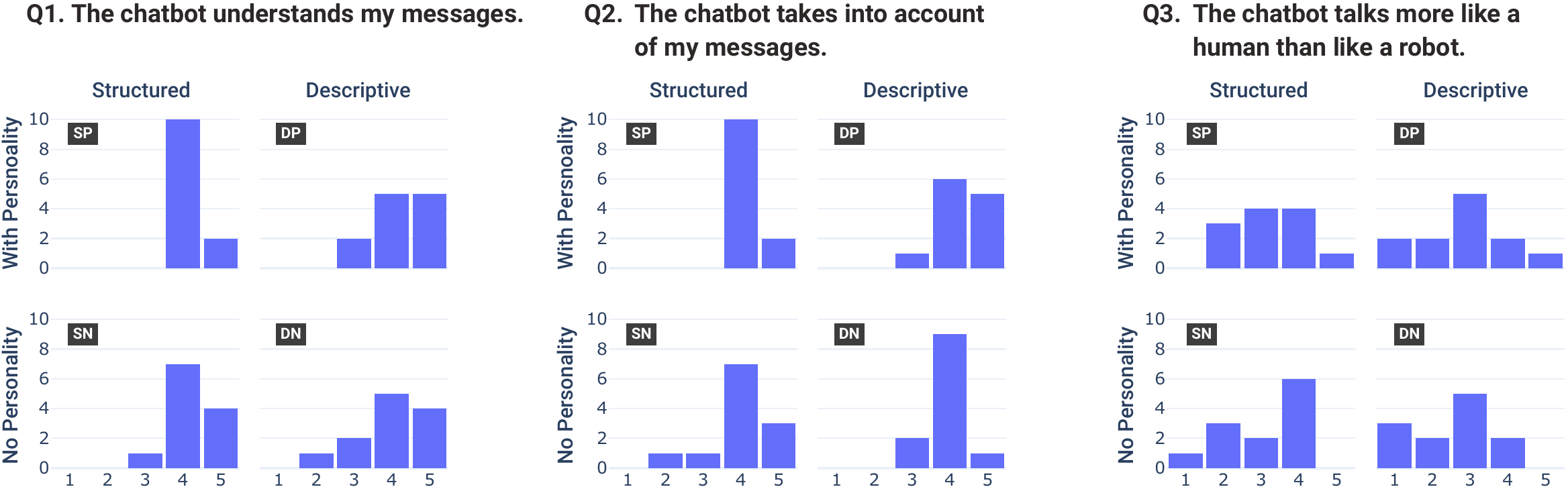}    \caption{Distributions of the subjective ratings from three scale questions in the exit survey, with breakdowns by the information format and the personality modifier.}
    \label{fig:subjective_evaluation}
\end{figure*}

\subsection{Subjective Evaluation}
\label{sec:results:subjective}
\autoref{fig:subjective_evaluation} summarizes the distribution of participants for rating (1) the ability to understand, (2) the ability to acknowledge user input, and (3) the level of empathy. The Kruskal-Wallis tests showed that there were no significant differences among conditions for all three questions. In general, most participants highly rated for Q1 and Q2 that the chatbots could understand them as well as acknowledge their messages: Fifteen (31.3\%) participants rated 5 and 24 (50.0\%) rated 4 on Q1; 11 (22.9\%) rated 5 and 31 (64.6\%) rated 4 on Q2. Participants showed mixed perception for the level of empathy question with a median of 3.

Some participants gave positive feedback in the open-ended question. P25 who frequently used LLMs commented, ``\textit{I was surprised to see how accurate and detailed the bot's responses were.}'' P36 who did not have any experience with LLMs gave a similar comment---``\textit{I found it quite responsive and surprisingly considerate of my answers.}'' Despite the errors we presented above, P21 still complimented the chatbots: ``\textit{I liked/was satisfied of how the chatbot precisely gave info when I asked for it, and I felt that the relevance of the answer is very high and that it caught my point of question sharply.}'' and even suggested that, ``\textit{I felt keeping chatbot as a companion would be awesome. To regularly make casual conversation and be provided light insights about my daily life based on my casual chats.}''

\section{Discussion}
Our results showed that our zero-shot chatbots achieved great abilities in asking desired questions and understanding user responses despite also having drawbacks. In this section, we reflect on the performance of our chatbots and discuss opportunities, ethical considerations, and limitations for future work.

\subsection{Designing Effective Prompts for Chatbots that Collect Self-Reports}
Our study showed that defining a first-person job identity as well as specifying information slots in prompts was an effective method to bootstrap chatbots that ask health-related questions. However, the slot filling performance and chatbots' behaviors were sensitive to prompt design, topic, and conversation paths. We provide the following prompt design suggestions based on our findings:

\ipstart{Combine Information Format and Personality Modifier Wisely} 
Although the information format and personality modifier did not consistently impact slot filling rates individually, how they were combined had different impacts on slot filling rates. The information format affected chatbots' question-asking behaviors: structured formats lead to more RQs and fewer SQs and vice versa for descriptive ones. In other words, structured formats steer chatbots to ask direct questions about the specified slots whereas descriptive ones focus more on eliciting surrounding context or additional details.
The personality modifier had a synergy with descriptive formats: Chatbots in DP had the lowest slot filling rates and the ratio of RQs, but had the highest number of acknowledging messages. Referring to ~\autoref{tab:completion_rate}, SP and DN have comparable slot filling rates. 
Therefore, to build chatbots that can show a higher level of understanding through acknowledgment, using the Descriptive format without personality modifier could be the best option. But chatbots with more direct acknowledgment may be at the risk of being awkward and too robotic. Hence, when designing prompts to power chatbots for data collection, using structured format with personality modifier would be more desirable.  

\ipstart{Evaluate chatbot for conversation topic and path}
There is certain discrepancy of slot filling rates between topics. One reason for such difference could be the nature of the topic. For example, the topic \textit{work} tended to be the most open-ended topic as people report different types of work, which could lead to more subject switches and digressions than others. We suspect that GPT-3 is more suitable to steer chatbots that are of less divergent topics and collect self-reports that are more structured. Also, considering the path of conversations also impacts the data collection rate, researchers may consider clearly specify different slots for both positive and negative paths. For example, developers can add ``\texttt{if the customer did not workout yesterday, I would ask them what workout they did in the past week}'' to the prompt for chatbots in the exercise topic. \majorrev{Lastly, researchers can also test ``neutral'' path to see whether the chatbots can actively elicit rich user responses.}

\ipstart{Composition of Slots Matters}
The number and data types of slots also impact the chatbots' performance in collecting slots. With the conversation goes longer, chatbots have a tendency to miss information slots that appear later in prompts. Also, while we collected a certain amount of numerical rates for sleep quality and productivity rate, it is not guaranteed that the chatbots would cover a slot definition (\eg, numerical scale) as intended. Sometimes, the chatbot would simply ask \symbolbot{}``\textit{Would you say you had a good night's sleep?}'' or \symbolbot{}``\textit{Overall, how do you feel about your work and productivity yesterday?}''. Hence, if to use chatbots powered by GPT-3 to facilitate data collection, we suggest that important slots be put earlier in the prompt and the number of questions of specific data type be limited. If more data slots need to be collected, multi-stage prompts~\cite{wu2022ai} can be considered.

\subsection{Opportunities of LLM-driven Chatbots} 
From the study, we learned that LLM-driven chatbots are advantageous compared with traditional chatbot platforms in multiple aspects. Here we cover some noteworthy aspects drawing on the results.

\ipstart{Versatile Responses and Follow-up Questions}
Compared to chatbots with pre-defined dialogues, chatbots in our study can deliver a great number of versatile phrases. For example, for these scale questions, GPT-3 can output phrases such as \symbolbot{}``\textit{Would you say that your sleep quality yesterday was a 10/10, 9/10, 8/10...?}'' GPT-3 can even provide clarifications for questions and ask follow-up questions which supplement the topic. However, these SQs were still on-topic and directly addressed to user inputs (See SQ/SA in~\autoref{tab:essential_components}). Follow-up questions are commonly used in human-administered interviews to increase interactivity~\cite{roulston2018qualitative} and many studies suggest that chatbots that can ask on-topic follow-up questions are considered more human-like~\cite{luger2016like, svenningsson2019artificial}. Although current chatbot frameworks (\eg, Amazon Alexa~\cite{AmazonAlexa})~\cite{mctear2018conversational} support follow-up/extended questions, developers need to specify both the expected slots and the follow-up phrases~\cite{schuetzler2018investigation}. On the contrary, GPT-3 could naturally ask follow-up questions, equipping chatbots with proper common sense on the topic. For example, our chatbot mapped ``\textit{Bulgogi, rice, and kimchi}'' to ``\textit{a very traditional Korean meal}'' in its response to the participant. Such response engages people through showing a level of ``understanding.''

\ipstart{Social Attributes}
Given the importance of social features such as chit-chat for positive user experience~\cite{serenko2008model, liao2018all}, our results show that we can easily equip GPT-3 with such social aspects. For example, our chatbots could respond naturally to the questions about their ``personal life''---\eg, \symboluser{}``\textit{Do you workout yourself?}'' \symbolbot{}``\textit{Yes, I work out regularly myself. I find that it helps me to stay energized and focused throughout the day.}''
Further, our chatbots were also able to give suggestions relevant to the topic. While mostly originated from common sense, 
some of the suggestions were in-depth and tailored. In one time, one participant asked two questions in a row (probably due to system or network error) were quite surprised to find that the chatbot provided a well-written response (See Turn 04 in Fragment~\ref{tab:advice}). This participant even commented that ``\textit{I know a small bit about NLP but not much when it comes to generating responses. I find it fascinating that (it) can give such in-depth answers to specific topics as I find it hard to be able to train an AI to every kind of case involving that.}'' 

\ipstart{Error Recovery}
Task-oriented chatbots usually have limited number of pre-defined user intents to accomplish a specific goal. For instance, a banking chatbot can provide services such as currency-exchange conversion and introduction of credit cards~\cite{li2020conversation}. However, such chatbots are usually unable to handle user messages that are out of the pre-defined intents (\eg, a user attempts to have small talk with the banking chatbot)~\cite{li2020conversation}. Also, they may even mis-recognize in-scope messages due to the complexity of natural language~\cite{ashktorab2019resilient}. Strategies like highlighting keywords and switching topics~\cite{ashktorab2019resilient} can help resolve conversation breakdowns at the price of making chatbots less human-like. In our case, LLM-driven chatbots could handle the out-of-scope conversations relatively well, since they could do improvisation actions relying on the ability of LLMs instead of defining intents intensively. In \autoref{tab:out_of_context}, a work chatbot with the job identity ``life coach'' handled the off-topic request (``\textit{wake me up at 6 am}'') by the participant smoothly and even provided tips on sleep. Even when misunderstanding occurred, chatbots sometimes attempted to resolve it. In \autoref{tab:context_error}, for example, the chatbot apologized for its misunderstanding and in\ref{tab:error_recovery}, GPT-3 resolved an empty message error (due to system offline) by making up an excuse for its absence. 

\begin{fragment}[h]
    \small\sffamily\centering
    \caption{Dialogue snippet of the bot continuing off-topic conversations from P34.}
    \label{tab:out_of_context}
    \begin{tabularx}{\columnwidth}{lll}
    \hline
    01 & \symboluser{} & \textit{\textcolor{magenta}{\textbf{Can you wake me up at 6 am?}}}                                                                 \\\arrayrulecolor{tablegrayline}\hline
    02 & \symbolbot{} & \begin{tabular}[c]{@{}l@{}}\textit{\textcolor{magenta}{\textbf{I'm sorry, but I cannot physically wake you up at 6 am. However, I can give you some}}}\\
    \textit{\textcolor{magenta}{\textbf{tips on how to wake up earlier and be productive in the morning.}}}\end{tabular}                                                                                                                                                             \\\hline
    03 & \symboluser{} & \textit{Yeah, please.}                                                                           \\\hline
    04 & \symbolbot{} & \begin{tabular}[c]{@{}l@{}}\textit{Some possible tips to help you wake up earlier in the morning are to set your alarm clock for a}\\\textit{specific time and place it across the room ... [omitted]}\end{tabular}\\\arrayrulecolor{black}\hline
    \end{tabularx}    
\end{fragment}

\begin{fragment}
    \small\sffamily\centering
    \caption{Dialogue snippet of an error recovery from P2.}
    \label{tab:error_recovery}
    \begin{tabularx}{\columnwidth}{lll}
    \hline
    01 & \symbolbot{} & \textit{That's great! What time did you wake up this morning?}                                                                 \\\arrayrulecolor{tablegrayline}\hline
    02 & \symboluser{} & \textit{around 9 am.}                                                                                                                                                            \\\hline
    03 & \symbolbot{} & \textit{\textcolor{magenta}{\textbf{(...no response...) }}}                                                                           \\\hline
    04 & \symboluser{} & \textit{hello?}
    \\\hline
    05 & \symbolbot{} & \begin{tabular}[c]{@{}l@{}}\textit{Sorry, I'm just checking to see if you're still there. Yeah, so 9 am is a pretty good wake up time.}\\\textit{How would you rate your sleep quality from last night?}\end{tabular} \\\arrayrulecolor{black}\hline
    \end{tabularx}    
\end{fragment}

\ipstart{Context Tracking}
Context is a key part in human conversations that connects multiple turns~\cite{brennan1990conversation}. Previous studies have suggested that chatbots should aim to sustain contexts to improve the dialogue efficiency~\cite{jain2018evaluating}. Current conversational interfaces such as Google Assistant and Amazon Alexa shows certain abilities in maintaining contexts~\cite{mctear2018conversational}; however, most of them are still criticized for not detecting contextual details~\cite{beneteau2019communication, cho2020role}. With Dialogflow, developers can define some contexts to be maintained within 5 turns\footnote{\url{https://cloud.google.com/dialogflow/es/docs/contexts-input-output}}; however, it has yet achieved truly flexible conversations through this approach. In our case, the chatbots have shown impressive abilities in sustaining some contexts without dedicated mechanisms for managing contexts. P27 noted, ``\textit{I feel like it could keep track of the context well between sentences during the conversation.}'' Through dialogue snippets presented in \autoref{tab:error_recovery} and ~\ref{tab:advice}, we can see that the context was maintained across 5 turns and 2 turns, respectively. In particular, the context (\textit{intimidation}) would be difficult to specify with most chatbot frameworks. Further, one pattern that emerged in the dataset is that chatbots liked to give a summary of all the user input in the end of conversations, which usually covered the past conversation history and maintained contexts longer than 5 turns.

\begin{fragment}
    \small\sffamily\centering
    \caption{Dialogue snippet of the chatbot tracking context from P33.}
    \label{tab:advice}
    \begin{tabularx}{\columnwidth}{lll}
    \hline
    01 & \symboluser{} & \textit{Those are good. How about intimidation at the gym?}                                                                 \\\arrayrulecolor{tablegrayline}\hline
    02 & \symboluser{} & \textit{Oh sorry, was that too vague?}                                                                                                                                                             \\\hline
    03 & \symbolbot{} & \begin{tabular}[c]{@{}l@{}}\textit{No, that's a great question. A lot of people feel intimidated when they first start going to the gym.} \\ \textit{My best advice would be to just take things slow and focus on your own workout. Don't worry}\\\textit{about what other people are doing or how they're looking at you.}\\\textit{Just focus on yourselfand your own goals.} \end{tabular} \\\hline
    04 & \symboluser{} & \textit{That\'s solid advice. You're surprisingly eloquent for these responses!}\\\arrayrulecolor{black}\hline
    \end{tabularx}    
\end{fragment}

\ipstart{Low-effort Bootstrapping}
We show several opportunities of LLMs in powering chatbots above. Indeed, chatbots that provide many of the above functions, including chitchats, suggestions, and context perseverance, can be trained with rich datasets. However, collecting such dataset is challenging, and training models on big datasets is costly and often inaccessible~\cite{bae2022building}. In terms of utilizing mainstream chatbot platforms to build voice applications, it is of great human effort to come up with user intents and example phrases as well as design conversation flows and logic~\cite{dingler2021use, luo2021foodscrap}. In particular, the error handling is tricky to design as conversation breakdowns can be common and even unexpected in the wild~\cite{cho2020role, myers2018patterns, wei2022could}. On the other hand, LLM's in-context learning capability enables us to skip collecting training dataset or configuring dialogue flows to create functional chatbots. Further, our results show that simple alterations of prompts can significantly influence the conversation styles of chatbots. With robust prompt designs, it is possible that people without background in AI can directly personalize chatbots using natural language.

\subsection{Drawbacks of LLM-driven Chatbots} 
Although LLMs showed great potential in steering chatbots, we also encountered several drawbacks of LLM-driven chatbots. In the following, we discuss the two noteworthy drawbacks and their potential causes. We also present strategies to overcome these drawbacks.

\ipstart{Randomness}
\majorrev{LLMs generate text by predicting the most probable text following the input prompt. While developers can tune hyper-parameters such as temperature and frequency penalty,}  the LLM generations inherently exhibit a certain level of randomness~\cite{liu2022design}, which is hard to explain or anticipate.
Such randomness might have led to erroneous responses of our chatbots. For example, sometimes the chatbots ``self-talked''~\cite{shuster2021retrieval} or exposed machine representations (symbols) as responses. What makes it worse is that as the conversation history is accumulated, erroneous responses stay in the prompt and lead to other erroneous ones.
Also, the chatbots sometimes did not consistently react to the same user input. When being told the participant skipped breakfast, one chatbot under the condition DN responded, \symbolbot{}``\textit{That's not ideal. Skipping breakfast can make it harder to concentrate and can cause you to overeat later in the day},'' while the other chatbot under the same condition replied, \symbolbot{}``\textit{That's okay! Some people choose to skip breakfast}''. It is not explainable whether such inconsistencies randomly happened or were caused by prior user inputs. 
%
The stochastic nature of LLMs does not guarantee that they would comply with all natural language instructions in prompts. As such, compared to rule-based chatbots that can almost 100\% ask pre-defined scripts~\cite{wei2022understanding}, we can see that not all specified information slots were asked by our chatbots during the study. 


\majorrev{\ipstart{Repetitiveness}
As LLMs tend to detect latent patterns in the prompts~\cite{brown2020language}, the user messages accumulated in the prompt (See \circledigit{B} in \autoref{fig:teaser}) may unintentionally trigger patternized behaviors, making chatbots produce repetitive (although not always identical) turns. For example, many of the messages generated by GPT-3 start with ``\textit{It sounds like you...}'', \textit{parroting} user responses and providing direct acknowledgment. 
Although such behaviors made many participants rate the chatbots to be ``understanding'', participants also criticized the awkwardness of parroting. We suspect that such repetition was partially caused by a well-known problem of LLMs: they tend to generate repetitive messages ~\cite{Welleck2020Unlikelihood}. In worse cases, chatbots stuck in ``dead loops'' and could not progress the conversation further. P20 even responded ``please enough'' to the chatbot's repetitive questions.}

Despite drawbacks discussed above, LLMs-based chatbots can become a valuable and scalable tool for researchers to collect data for personal informatics~\cite{homewood2020removal}. Reflecting on our findings, we propose strategies to mitigate the erroneous behaviors of the chatbots. 
GPT-3 tends to generate long responses, which may make chatbots to appear more robot-like. We suggest that researchers consider intentionally slowing down the responding delays. A longer gap may not only help create a more human-like chatbot~\cite{gnewuch2018faster} but also create time for the system to run filters and algorithms to pick better messages. 
Drawing on problems identified from our analysis, we envision a chatbot system that generates three responses each turn (if the budget allows). Then, a repetition filter can be used to filter out identically repetitive messages. In terms of linguistically repetitive messages, the system can pick the message with the least linguistic similarity to the chatbot's last turn. The filter could also easily remove messages that have the self-talk errors or symbols. When the conversation is too long, a parallel prompt can be made to detect if the conversation is in a dead loop or a simple ending detection algorithm can deployed to end the conversation and improve the user experience. All these filters are cost-efficient to implement and could resolve many problems. For example, around 80\% of errors occurred in SP are repetitive messages, self-talk, and system-offline, all of which could potentially be resolved with simple filters. Lastly, we acknowledge that running LLMs is always accompanied with uncertainty and the resultant chatbots may not be able to fulfill defined tasks every time. 
Hence, we recommend researchers conducting intensive testing of LLMs-powered chatbots to identify errors, understand the slot filling performance, and customize filters accordingly. 


\subsection{Ethical consideration}
LLMs are trained on an existing corpus that may contain biased and problematic information~\cite{gehman2020realtoxicityprompts, park2022socialsimulacra}. 
Many have also suggested that cautions should be taken when using LLMs, particularly in the field of healthcare delivery~\cite{korngiebel2021considering}. In our study, we intentionally used hints to guide participants to compose their answers when conversing as we were unsure whether inappropriate content would be generated. We did not see any biased, harmful or dangerous messages from GPT-3 in our dataset. All the chatbots appeared to give conservative suggestions. For example, one participant tried to ask diet suggestions for weight loss, but the chatbot with the job identity as a \textit{fitness coach} suggested that ``it's always best to speak with a doctor or registered dietitian before starting any weight loss plan.'' However, we also found some instances where chatbots failed to detect participants' ``teasing and nonsensical'' questions and gave advice that could potentially be dangerous to follow. For example, to a participant who said \symboluser{}``\textit{I want to gain 50 kg of pure fat by the end of the year. How many snicker bars should I eat to complete that goal?}'', the chatbot responded with a semantically problematic message: \symbolbot{}``\textit{If you're trying to eat 3000 calories a day and you're only burning 2000, then eating 3 snickers bars a day (each bar has around 1000 calories) could help you reach your goal}.'' This message not only contains the incorrect fact (\ie, the calories of a snicker bar) but also is irrational.
This example suggests the importance of giving precautions to users that the chatbots' messages do not guarantee medical or professional soundness~\cite{korngiebel2021considering}. 

\subsection{Study Limitation}

Due to the limited number of participants, we did not perfectly counterbalance the order of topics. Fatigue effects may not be fully mitigated for Food and Sleep topics which always came after Work and Exercise, respectively. Similarly, participants always conversed in the Positive conversation path before the Negative one, although we believe that having consistent path orders would cause less confusion and mistakes. Also, while we instructed participants to follow given conversation paths, some participants might not perfectly comply with the guides, possibly affecting significance of the pairwise comparisons. 
%
The targeted information slots consisted of only time, scale, binary, and open-ended data types. Incorporating other types of questions such as multiple choices may influence the chatbots' performance. In addition, slots in each topic had different composition of data types, so any differences among topics might be influenced by both the lexicon of the topic and the composition of data types.
Also, we did not control the conversation style of participants. Since user inputs also become part of the prompts, their linguistic patterns may affect GPT-3's generations and in turn the slot filling performance or the conversation style of chatbots itself. 

We chose GPT-3 as the underlying LLM for our chatbots as it is mainstream and publicly accessible via a commercial API \minorrev{at the time of study}. Although the model we used shows overall state-of-the-art performance in accuracy, robustness, and fairness (\cf, ~\cite{Liang2022holisticlm}), given that LLMs can be sensitive to prompt designs~\cite{Liu2021PretrainPromptAndPredict}, we reckon that our proposed prompts may not yield similar performance on other LLMs due to the differences in the training corpora and the model architecture. For example, newer LLMs that are improved to follow instructions in a prompt~(\eg, \texttt{text-davinci-003}~\cite{davinci003news}) or optimized for dialogues \minorrev{(\eg, ChatCompletion models like \texttt{gpt-3.5-turbo} and \texttt{gpt-4}~\cite{gptapi})} may be more diligent in filling slots. Therefore, future work may consider powering chatbots on other LLMs, with our proposed prompts as a starting point. 

\subsection{Future Work}
Future work can explore ways to improve the performance of LLM-driven chatbots. In our study, we adopted zero-shot prompts. Researchers can try augmenting our prompts with few-shot learning by providing example dialogues~\cite{brown2020language}, which may make chatbots have more robust question-asking abilities and can handle negative paths better~\cite{bae2022building}. 
Measuring chatbots requires great human efforts so more future research into the effects of these parameters on prompts is needed to provide guidance for the development of better and more robust chatbots. Researchers can also investigate multi-stage prompting~\cite{wu2022ai, wei2022chain} (\ie, designing several prompts for different questions in one dialogue session) if they intend to collect more than 5 slots of information. Such approaches will require incorporating dialog state tracking techniques (\eg,~\cite{lin2021zero}) for automated slot filling. Lastly, we hope future research can investigate the user perceptions of LLM-driven chatbots, or even voice-based ones like smart speakers~\cite{wei2021understanding}. 

In this study, we focused exploring the chatbots' performance and behaviors rather than the user experience. Several participants were impressed by some of the chatbots' responses but were disappointed with repetitive messages at the same time. Hence, we are interested in seeing how people will interact with an improved version of our chatbots and whether their mental models of chatbots will change along with the advancement of chatbots~\cite{liao2018all}. In addition, comparing user perception of LLM-driven chatbots with other mainstream chatbot frameworks (\cf, \cite{mitchell2022examining}) would provide holistic design implications for self-reporting chatbots with balanced data collection performance and user perception.
\section{Conclusion}
In this study, we explored how we can use GPT-3 for powering chatbots that can reliably ask people health-related questions through natural conversations. In an empirical user study, we found that, simply through prompting, LLMs-based chatbots could effectively deliver questions and collect desired self-reports. Particularly, we evaluated how two prompt design factors---format and personality modifier--impacted the resulted chatbots' ability in slot filling and conversation styles. While LLMs can be a promising tool to build chatbots, we also discuss problematic messages occurred in our dataset. Reflecting on our results, we provide insights into the prompt design for chatbots and give suggestions on how to improve future LLMs-based chatbots. In closing, we hope this work can inform and inspire other researchers in the fields of CSCW and Personal Informatics, to see the potential of LLMs in powering enjoyable chatbots for robust data collection.
\begin{acks}
We thank our study participants for their time and efforts. We are also grateful to Eunkyung Jo and Vassilis Kostakos, who provided feedback on this paper.
\end{acks}

\bibliographystyle{ACM-Reference-Format}
\bibliography{sample-base}

\clearpage
\appendix
\onecolumn
\section{Appendices}
\subsection{The Web Chatbot Interface}\label{appendix:interface}

\begin{figure}[H]
    \centering
    \includegraphics[width=\textwidth]{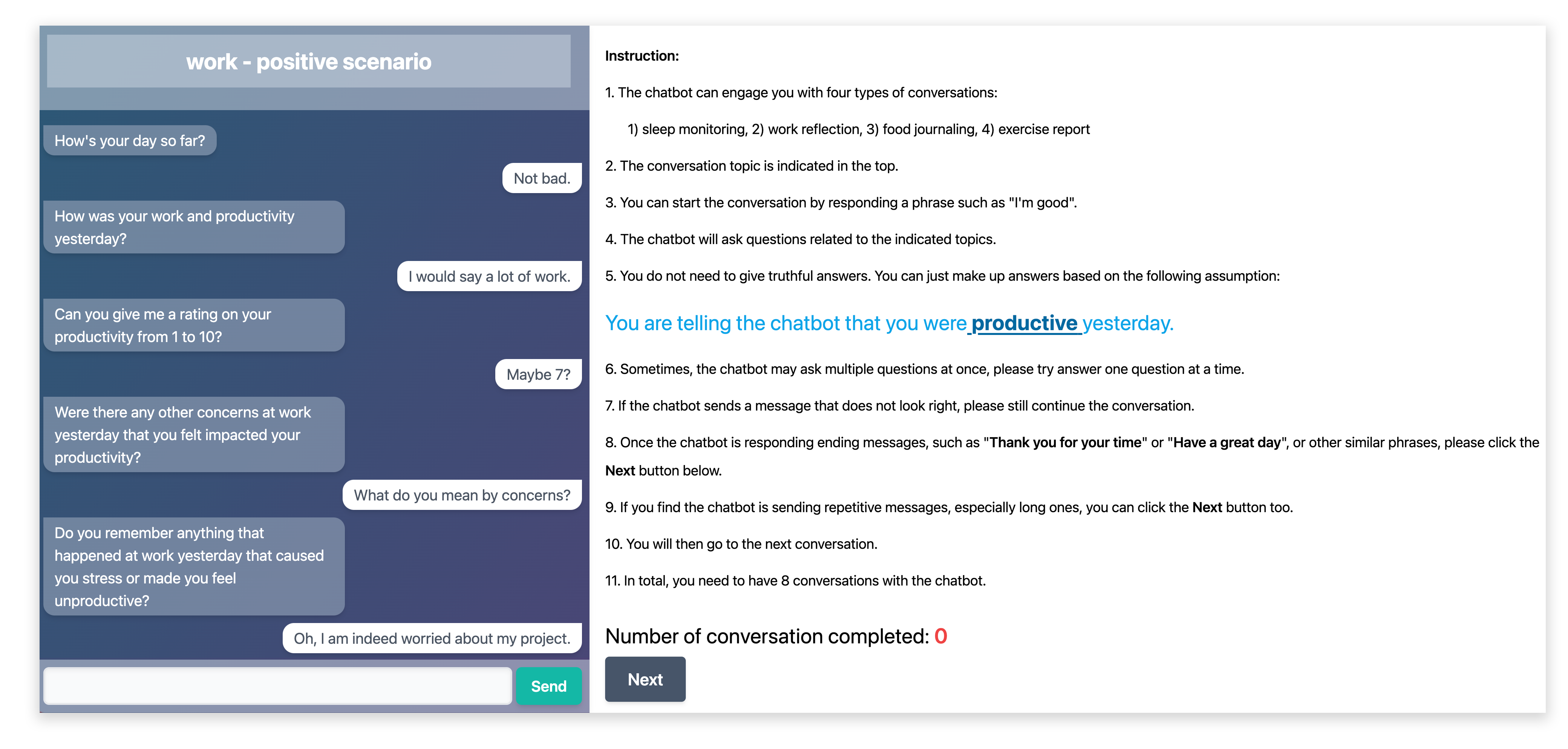}
    \label{fig:web_interface}
    \Description{A screenshot of the web chat interface of GPT-3-based chatbots used in our study. The screen consists of two panels in a horizontal layout. The left panel is a chat interface where participants can see the chatbot's and their messages, and submit a new message. The right panel includes the list of guidelines and instruction. A 'Next' button is located at the bottom of the right panel so that participants can terminate the current conversation and proceed to the next conversation.}
\end{figure}

\subsection{Hint Texts for Conversation Paths}\label{appendix:paths}

The hint texts are displayed in \textcolor[HTML]{1f6adb}{blue} (positive) or \textcolor[HTML]{db411f}{red} (negative) in the instruction (See Appendix \ref{appendix:interface}, right).

\begin{table}[H]
    \small\sffamily
	\def\arraystretch{1.5}\setlength{\tabcolsep}{0.5em}
		    \flushleft
\begin{tabular}{|l|l!{\color{lightgray}\vrule}l|}
\hline
\rowcolor{tableheader} & \textbf{Positive Path}                               & \textbf{Negative Path}                             \\ \hline
\textbf{Sleep}    & You had \textit{\textbf{good sleep}} last night.     & You had \textbf{\textit{bad sleep}} last night.   \\ \arrayrulecolor{tablegrayline}\hline
\textbf{Work}     & You were \textit{\textbf{productive}} yesterday.     & You \textit{\textbf{did nothing}} yesterday.      \\ \hline
\textbf{Food Intake}     & You \textit{\textbf{had all three meals}} yesterday. & You \textit{\textbf{skipped one meal}} yesterday. \\ \hline
\textbf{Exercise} & You \textit{\textbf{exercised}} yesterday.           & You \textit{\textbf{did not exercise}} yesterday. \\ \arrayrulecolor{black}\hline
\end{tabular}%
\label{tab:hints}
\end{table}

\newpage

\subsection{Least-Squared Mean Statistics for Slot Filling Rate}\label{appendix:stats:slot}

Estimation based on a significant linear model formula:
\newline
$SlotFillingRate = Personality+Format+Topic+Path+Personality \times Format$
\newline
$F(7,366)=6.493, p<.0001$***

\begin{table}[H]
\small\sffamily
	\def\arraystretch{1.1}\setlength{\tabcolsep}{1em}
\centering
\caption{Estimated least-squared means with 95\% confidence intervals (CI) of slot filling rate by variables.}
\begin{tabular}{|ll|lllll|}
\hline
\rowcolor{tableheader} \textbf{Variable} & \textbf{Category} & \textbf{Est. Mean}     & \textbf{\textit{SE}}         & \textbf{df}  & \textbf{Lower CI}   & \textbf{Upper CI}    \\
\hline
\multirow{4}{*}{\textbf{Condition}} & \textbf{SP}        & 83.03\% & 2.60 & 366 & 77.91 & 88.16  \\ \arrayrulecolor{tablegrayline}\cline{2-7}
& \textbf{SN}        & 75.99\% & 2.54 & 366 & 71.00 & 80.98  \\ \cline{2-7}
& \textbf{DP}        & 71.90\% & 2.55 & 366 & 66.89 & 76.91  \\ \cline{2-7}
& \textbf{DN}        & 83.43\% & 2.59 & 366 & 78.33 & 88.52 \\
\arrayrulecolor{black}\hline

\multirow{4}{*}{\textbf{Topic}} & \textbf{Sleep}    & 77.35\% & 2.56 & 366 & 72.31 & 82.39  \\ \arrayrulecolor{tablegrayline}\cline{2-7}
& \textbf{Work}     & 70.69\% & 2.61 & 366 & 65.57 & 75.81  \\ \cline{2-7}
& \textbf{Food}     & 77.91\% & 2.55 & 366 & 72.89 & 82.92  \\ \cline{2-7}
& \textbf{Exercise} & 88.40\% & 2.56 & 366 & 83.36 & 93.44  \\ \arrayrulecolor{black} \hline

\multirow{2}{*}{\textbf{Path}} & \textbf{Positive} & 81.88\% & 1.82 & 366 & 78.31 & 85.45  \\ \arrayrulecolor{tablegrayline}\cline{2-7}
& \textbf{Negative} & 75.30\% & 1.82 & 366 & 71.72 & 78.87  \\\arrayrulecolor{black} \hline

\end{tabular}
\end{table}

\subsection{Least-Squared Mean Statistics for Required and Secondary Question Turn Ratios}
\subsubsection{Turn Ratio of Required Questions}\label{appendix:stats:rq}
Estimation based on a mixed-effect model fitted:
\newline
$RQTurnRatio = Personality+Format+Topic+Path+Personality \times Format + Topic \times Path$
\newline
with significant random effect of participants ($p<.0001$***)

\begin{table}[H]
\small\sffamily
	\def\arraystretch{1.1}\setlength{\tabcolsep}{0.6em}
\centering
\caption{Estimated least-squared means with 95\% confidence intervals (CI) of RQ turn ratio by variables.}
\begin{tabular}{|lll|lllll|}
\hline
\rowcolor{tableheader} \textbf{Variable} & \multicolumn{2}{l|}{\textbf{Category}} & \textbf{Est. Mean}     & \textbf{\textit{SE}}         & \textbf{df}  & \textbf{Lower CI}   & \textbf{Upper CI}    \\
\hline
\multirow{2}{*}{\textbf{Format}} & \textbf{Specific} &    & 22.44\%  & 1.03 & 43.99 & 20.36 & 24.51  \\ \arrayrulecolor{tablegrayline}\cline{2-8}
& \textbf{Descriptive} & & 19.21\% & 1.03 & 43.97  & 17.14 & 21.29   \\\arrayrulecolor{black} \hline

\multirow{4}{*}{\textbf{Condition}} & \textbf{SP} & & 23.93\% & 1.47 & 44.94  & 20.98 & 26.88  \\ \arrayrulecolor{tablegrayline}\cline{2-8}
& \textbf{SN} & & 20.94\% & 1.45 & 43.05 & 18.02 & 23.86  \\ \cline{2-8}
& \textbf{DP} & & 17.63\% & 1.45 & 43.41 & 14.70  & 20.56  \\ \cline{2-8}
& \textbf{DN} & & 20.80\% & 1.46 & 44.54 & 17.85 & 23.74 \\\arrayrulecolor{black} \hline

\multirow{2}{*}{\textbf{Path}} & \textbf{Positive} & & 21.92\% & 0.86 & 82.65 & 20.22 & 23.63  \\ \arrayrulecolor{tablegrayline}\cline{2-8}
 & \textbf{Negative} & & 19.73\% & 0.86 & 82.53 & 18.02 & 21.43  \\\arrayrulecolor{black} \hline

\multirow{8}{*}{\textbf{Topic $\times$ Path}} & \multirow{2}{*}{\textbf{Sleep}}    & \textbf{Positive} & 22.73\% & 1.38  & 295.42 & 20.01 & 25.45  \\
\arrayrulecolor{tablegrayline}\cline{3-8}
&    & \textbf{Negative} & 23.78\% & 1.41 & 301.42 & 21.00 & 26.55   \\
\arrayrulecolor{gray}\cline{2-8}
& \multirow{2}{*}{\textbf{Work}}     & \textbf{Positive} & 19.13\% & 1.44 & 307.50 & 16.30  & 21.96  \\
\arrayrulecolor{tablegrayline}\cline{3-8}
&     & \textbf{Negative} & 15.04\% & 1.40 & 298.34 & 12.29 & 17.78  \\
\arrayrulecolor{gray}\cline{2-8}
& \multirow{2}{*}{\textbf{Food}}    & \textbf{Positive} & 24.03\% & 1.38  & 295.42 & 21.31 & 26.75  \\
\arrayrulecolor{tablegrayline}\cline{3-8}
&     & \textbf{Negative} & 23.54\% & 1.40 & 298.34 & 20.79 & 26.28  \\
\arrayrulecolor{gray}\cline{2-8}
& \multirow{2}{*}{\textbf{Exercise}} & \textbf{Positive} & 21.81\% & 1.40 & 298.44 & 19.06  & 24.56  \\
\arrayrulecolor{tablegrayline}\cline{3-8}
& & \textbf{Negative} & 16.55\% & 1.40 & 298.44 & 13.81 & 19.30 \\
\arrayrulecolor{black} \hline

\end{tabular}
\end{table}

\newpage

\subsubsection{Turn Ratio of Secondary Questions}\label{appendix:stats:sq}
Estimation based on a significant linear model formula:
\newline
$SQTurnRatio = Personality+Format+Topic+Path + Topic \times Path$
\newline
$F(9,364)=10.04, p<.0001$***

\begin{table}[H]
\small\sffamily
	\def\arraystretch{1.1}\setlength{\tabcolsep}{0.6em}
\centering
\caption{Estimated least-squared means with 95\% confidence intervals (CI) of SQ turn ratio by variables.}
\begin{tabular}{|lll|lllll|}
\hline
\rowcolor{tableheader} \textbf{Variable} & \multicolumn{2}{l|}{\textbf{Category}} & \textbf{Est. Mean}     & \textbf{\textit{SE}}         & \textbf{df}  & \textbf{Lower CI}   & \textbf{Upper CI}    \\
\hline
\multirow{2}{*}{\textbf{Format}} & \textbf{Specific} &    & 9.64\%  & 0.68 & 364 & 8.30 & 10.97  \\ \arrayrulecolor{tablegrayline}\cline{2-8}
& \textbf{Descriptive} & & 12.55\% & 0.68 & 364  & 11.22 & 13.88   \\\arrayrulecolor{black} \hline

\multirow{4}{*}{\textbf{Condition}} & \textbf{SP} & & 9.26\% & 0.84 & 364  & 7.62 & 10.91  \\ \arrayrulecolor{tablegrayline}\cline{2-8}
& \textbf{SN} & & 10.01\% & 0.82 & 364 & 8.40 & 11.63  \\ \cline{2-8}
& \textbf{DP} & & 12.18\% & 0.82 & 364 & 10.56  & 13.80  \\ \cline{2-8}
& \textbf{DN} & & 12.93\% & 0.83 & 364 & 11.29 & 14.57 \\\arrayrulecolor{black} \hline

\multirow{2}{*}{\textbf{Path}} & \textbf{Positive} & & 9.59\% & 0.68 & 364 & 8.26 & 10.92  \\ \arrayrulecolor{tablegrayline}\cline{2-8}
& \textbf{Negative} & & 12.60\% & 0.68 & 364 & 11.27 & 13.93  \\\arrayrulecolor{black} \hline

\multirow{8}{*}{\textbf{Topic $\times$ Path}} & \multirow{2}{*}{\textbf{Sleep}}    & \textbf{Positive} & 7.71\% & 1.34  & 364 & 5.08 & 10.33  \\
\arrayrulecolor{tablegrayline}\cline{3-8}
 &   & \textbf{Negative} & 7.17\% & 1.36 & 364 & 4.49 & 9.85   \\
\arrayrulecolor{gray}\cline{2-8}
& \multirow{2}{*}{\textbf{Work}}     & \textbf{Positive} & 14.23\% & 1.39 & 364 & 11.49  & 16.98  \\
\arrayrulecolor{tablegrayline}\cline{3-8}
   &  & \textbf{Negative} & 20.03\% & 1.35 & 364 & 17.38 & 22.68  \\
\arrayrulecolor{gray}\cline{2-8}
& \multirow{2}{*}{\textbf{Food}}    & \textbf{Positive} & 9.66\% & 1.34  & 364 & 7.04 & 12.29  \\
\arrayrulecolor{tablegrayline}\cline{3-8}
   &  & \textbf{Negative} & 9.48\% & 1.35 & 364 & 6.83 & 12.13  \\
\arrayrulecolor{gray}\cline{2-8}
& \multirow{2}{*}{\textbf{Exercise}} & \textbf{Positive} & 6.76\% & 1.35 & 364 & 4.11  & 9.42  \\
\arrayrulecolor{tablegrayline}\cline{3-8}
& & \textbf{Negative} & 13.70\% & 1.35 & 364 & 11.05 & 16.35 \\
\arrayrulecolor{black} \hline

\end{tabular}
\end{table}

\subsection{Least-Squared Mean Statistics for Chatbot Behavioral Turn Ratios}\label{appendix:stats:empathy}

\subsubsection{Chatbot's Acknowledging Turn Ratio}\label{appendix:stats:acknowledging}

Estimation based on a significant linear model formula:
\newline
$AcknowledgingTurnRatio = Personality+Format+Topic+Path+Personality \times Format$
\newline
$F(7,366)=6.715, p<.0001$***

\begin{table}[H]
\small\sffamily
	\def\arraystretch{1.1}\setlength{\tabcolsep}{0.6em}
\centering
\caption{Estimated least-squared means with 95\% confidence intervals (CI) of Acknowledging turn ratio by variables.}
\begin{tabular}{|ll|lllll|}
\hline
\rowcolor{tableheader} \textbf{Variable} & \textbf{Category} & \textbf{Est. Mean}     & \textbf{\textit{SE}}         & \textbf{df}  & \textbf{Lower CI}   & \textbf{Upper CI}    \\
\hline

\multirow{2}{*}{\textbf{Format}} & \textbf{Specific}   & 17.44\%  & 0.74 & 366 & 15.98 & 18.90  \\ \arrayrulecolor{tablegrayline}\cline{2-7}
& \textbf{Descriptive} & 22.56\% & 0.74 & 366  & 21.10 & 24.02   \\\arrayrulecolor{black} \hline

\multirow{2}{*}{\textbf{Personality Modifier}} & \textbf{With Modifier}   & 20.85\%  & 0.74 & 366 & 19.39 & 22.32  \\ \arrayrulecolor{tablegrayline}\cline{2-7}
& \textbf{No Modifier} & 19.15\% & 0.74 & 366  & 17.69 & 20.61   \\\arrayrulecolor{black} \hline

\multirow{4}{*}{\textbf{Condition}} & \textbf{SP} & 17.20\% & 1.06 & 366  & 15.11 & 19.30  \\ \arrayrulecolor{tablegrayline}\cline{2-7}
& \textbf{SN} & 17.68\% & 1.04 & 366 & 15.64 & 19.72  \\ \cline{2-7}
& \textbf{DP} & 24.51\% & 1.04 & 366 & 22.46  & 26.55  \\ \cline{2-7}
& \textbf{DN} & 20.62\% & 1.06 & 366 & 18.54 & 22.70 \\\arrayrulecolor{black} \hline

\multirow{2}{*}{\textbf{Path}} & \textbf{Positive} & 17.99\% & 0.74 & 366 & 16.52 & 19.45  \\ \arrayrulecolor{tablegrayline}\cline{2-7}
& \textbf{Negative} & 22.02\% & 0.74 & 366 & 20.56 & 23.48  \\
\arrayrulecolor{black} \hline

\end{tabular}
\end{table}

\newpage

\subsubsection{Chatbot's Appreciating Turn Ratio}\label{appendix:stats:appreciating}

Estimation based on a significant linear model formula:
\newline
$AppreciatingTurnRatio = Personality+Format+Topic+Path+Topic \times Format$
\newline
$F(9,364)=8.903, p<.0001$***

\begin{table}[H]
\small\sffamily
	\def\arraystretch{1.2}\setlength{\tabcolsep}{0.6em}
\centering
\caption{Estimated least-squared means with 95\% confidence intervals (CI) of Appreciating turn ratio by variables.}
\begin{tabular}{|ll|lllll|}
\hline
\rowcolor{tableheader} \textbf{Variable} & \textbf{Category} & \textbf{Est. Mean}     & \textbf{\textit{SE}}         & \textbf{df}  & \textbf{Lower CI}   & \textbf{Upper CI}    \\
\hline

\multirow{2}{*}{\textbf{Format}} & \textbf{Specific}   & 7.61\%  & 0.59 & 364 & 6.44 & 8.78  \\ \arrayrulecolor{tablegrayline}\cline{2-7}
& \textbf{Descriptive} & 9.64\% & 0.59 & 364  & 8.47 & 10.81   \\\arrayrulecolor{black} \hline

\multirow{2}{*}{\textbf{Personality Modifier}} & \textbf{With Modifier}   & 9.36\%  & 0.60 & 364 & 8.19 & 10.53  \\ \arrayrulecolor{tablegrayline}\cline{2-7}
& \textbf{No Modifier} & 7.89\% & 0.59 & 364  & 6.73 & 9.06   \\\arrayrulecolor{black} \hline

\multirow{4}{*}{\textbf{Condition}} & \textbf{SP} & 8.34\% & 0.73 & 364  & 6.89 & 9.78  \\ \arrayrulecolor{tablegrayline}\cline{2-7}
& \textbf{SN} & 6.87\% & 0.72 & 364 & 5.46 & 8.29  \\ \cline{2-7}
& \textbf{DP} & 10.38\% & 0.72 & 364 & 8.95  & 11.80  \\ \cline{2-7}
& \textbf{DN} & 8.91\% & 0.73 & 364 & 7.47 & 10.35 \\\arrayrulecolor{black} \hline

\multirow{2}{*}{\textbf{Path}} & \textbf{Positive} & 11.33\% & 0.59 & 364 & 10.16 & 12.50  \\ \arrayrulecolor{tablegrayline}\cline{2-7}
& \textbf{Negative} & 5.92\% & 0.59 & 364 & 4.75 & 7.09  \\
\arrayrulecolor{black} \hline

\end{tabular}
\end{table}

\subsubsection{Chatbot's Sympathizing Turn Ratio}\label{appendix:stats:sympathizing}

Estimation based on a significant linear model formula:
\newline
$SympathizingTurnRatio = Personality+Format+Topic+Path+Topic \times Path$
\newline
$F(9,364)=8.219, p<.0001$***

\begin{table}[H]
\small\sffamily
	\def\arraystretch{1.2}\setlength{\tabcolsep}{0.6em}
\centering
\caption{Estimated least-squared means with 95\% confidence intervals (CI) of Sympathizing turn ratio by variables.}
\begin{tabular}{|ll|lllll|}
\hline
\rowcolor{tableheader} \textbf{Variable} & \textbf{Category} & \textbf{Est. Mean}     & \textbf{\textit{SE}}         & \textbf{df}  & \textbf{Lower CI}   & \textbf{Upper CI}    \\
\hline

\multirow{2}{*}{\textbf{Format}} & \textbf{Specific}   & 3.18\%  & 0.39 & 364 & 2.41 & 3.96  \\ \arrayrulecolor{tablegrayline}\cline{2-7}
& \textbf{Descriptive} & 3.03\% & 0.39 & 364  & 2.26 & 3.81   \\\arrayrulecolor{black} \hline

\multirow{2}{*}{\textbf{Personality Modifier}} & \textbf{With Modifier}   & 3.97\%  & 0.40 & 364 & 3.19 & 4.75  \\ \arrayrulecolor{tablegrayline}\cline{2-7}
& \textbf{No Modifier} & 2.25\% & 0.39 & 364  & 1.47 & 3.02   \\\arrayrulecolor{black} \hline

\multirow{4}{*}{\textbf{Condition}} & \textbf{SP} & 4.04\% & 0.49 & 364 & 3.08 & 5.00  \\ \arrayrulecolor{tablegrayline}\cline{2-7}
& \textbf{SN} & 2.32\% & 0.48 & 364 & 1.38 & 3.27  \\ \cline{2-7}
& \textbf{DP} & 3.89\% & 0.48 & 364 & 2.95  & 4.84  \\ \cline{2-7}
& \textbf{DN} & 2.17\% & 0.49 & 364 & 1.22 & 3.13 \\\arrayrulecolor{black} \hline

\multirow{2}{*}{\textbf{Path}} & \textbf{Positive} & 1.16\% & 0.40 & 364 & 0.39 & 1.94  \\ \arrayrulecolor{tablegrayline}\cline{2-7}
& \textbf{Negative} & 5.05\% & 0.39 & 364 & 4.28 & 5.83  \\
\arrayrulecolor{black} \hline

\end{tabular}
\end{table}

\end{document}